\documentclass[reprint,prb,aps]{revtex4-1}

\usepackage{graphicx}
\usepackage{dcolumn}
\usepackage{bm}
\usepackage{amsmath}
\usepackage{epstopdf}

\newlength{\figwidth}
\newlength{\figwidthb}
\newcommand{\CIO}{Ca$_{5}$Ir$_3$O$_{12}$\,\,}

\begin{document}

\title{Probing electronic excitations in iridates with resonant inelastic x-ray scattering and emission spectroscopy techniques}
\author{Young-June Kim}
\email{yjkim@physics.utoronto.ca} \affiliation{Department of
Physics, University of Toronto, Toronto, Ontario M5S~1A7, Canada}
\author{J. P. Clancy}
\affiliation{Department of Physics, University of Toronto, Toronto,
Ontario M5S~1A7, Canada}
\author{H. Gretarsson}
\affiliation{Department of Physics, University of Toronto, Toronto,
Ontario M5S~1A7, Canada}
\author{G. Cao}
\affiliation{Department of Physics, University of Colorado Boulder, Boulder, Colorado 80309, USA}
\author{Yogesh Singh}
\affiliation{Indian Institute of Science Education and Research Mohali, Sector 81, SAS Nagar, Manauli PO 140306, India}
\author{Jungho Kim}
\affiliation{Advanced Photon Source,
Argonne National Laboratory, Argonne, Illinois 60439, USA}
\author{M. H. Upton}
\affiliation{Advanced Photon Source,
Argonne National Laboratory, Argonne, Illinois 60439, USA}
\author{D. Casa}
\affiliation{Advanced Photon Source,
Argonne National Laboratory, Argonne, Illinois 60439, USA}
\author{T. Gog}
\affiliation{Advanced Photon Source,
Argonne National Laboratory, Argonne, Illinois 60439, USA}

\date{\today}

\begin{abstract}
We report a comprehensive resonant inelastic x-ray scattering (RIXS) study of various iridate materials focusing on core-level excitations and transitions between crystal-field split d-levels. The 2p core hole created at the Ir $L_3$ absorption edge has a very short lifetime giving rise to a broad absorption width ($\sim 5$~eV). This absorption linewidth broadening can be overcome by studying the resonant x-ray emission spectroscopy (RXES) map, which is a two-dimensional intensity map of the Ir L$\alpha_2$ emission obtained with high energy-resolution monochromator and analyzer. By limiting the emitted photon energy to a narrow range, one can obtain x-ray absorption spectra in the high energy-resolution fluorescence detection (HERFD) mode, while one can also simulate quasi-$M_4$-edge absorption spectra by integrating over incident photon energies. Both methods improve the absorption line width significantly, allowing detailed studies of unoccupied electronic structure in iridates and other $5d$ transition metal compounds. On the other hand, the short lifetime of the 2p core hole benefits the study of excitations of valence electrons. We show that the incident energy dependence of the RIXS spectra for $d-d$ transitions is simple to understand due to the short core-hole lifetime, which validates ultra-short core-hole lifetime approximation used widely in theoretical calculations. We compared $d-d$ excitations in various iridates and found that the excitations between the t$_{2g}$ and e$_g$ states share many similarities among different materials. However, the RIXS spectra due to the transitions between the spin-orbit-split t$_{2g}$ levels vary widely depending on the oxidation state and electronic bandwidths.
\end{abstract}

\pacs{} \maketitle

\section{introduction}

\begin{figure}[!ht]
\begin{center}
\includegraphics[angle=0,width=3.1in]{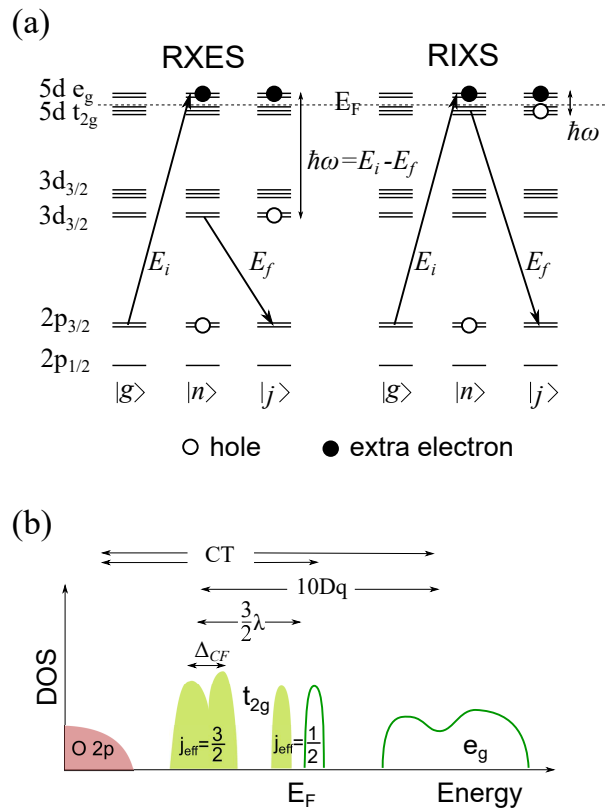}
\end{center}
\caption{(Color online) (a) Schematic diagram describing resonant x-ray emission spectroscopy (RXES) and resonant inelastic x-ray scattering (RIXS) processes. Both are  photon-in photon-out spectroscopy methods that measure the energy loss of the photon. In the case of RXES, the energy loss corresponds to transitions between deep core levels and unoccupied states just above the Fermi level. RIXS measures the valence electron's excitation across the Fermi energy. $|g\rangle$, $|n\rangle$, and $|j\rangle$ denote electron's ground, intermediate, and final states, respectively. We reserve $i$ and $f$ subscripts to denote photon initial and final energies. (b) Schematic representation of the density of states for an Ir$^{4+}$ system. Here we assume significant spin-orbit coupling $\lambda$ and moderate electron correlation to open a Mott gap. 10Dq represents the octahedral crystal field splitting, and $\Delta_{CF}$ is non-cubic crystal field splitting (e.g., trigonal or tetragonal distortion). The charge-transfer (CT) energy is assumed to be much larger than any other energy scale. Shaded area represents filled states.}\label{fig:schematic}
\end{figure}

Resonant inelastic x-ray scattering (RIXS) has made significant contributions to understanding the physics of cuprates and iridates in the past decade.
RIXS is a second-order scattering process which can be used to probe
elementary excitations involving spin, orbital, charge, and lattice degrees of freedom \cite{Kotani2001,Rueff2010,Ament2011}. When the incident photon energy is tuned to the relevant absorption edge, the scattering cross-section of these elementary excitations receives a large resonant enhancement. Past investigations have been mostly focused on core level excitations, which is often called resonant x-ray emission spectroscopy (RXES) to emphasize the ``emission" nature of excitations \cite{Schulke_book,deGroot_book}. However, in recent years, with improved energy resolution, RIXS has been widely used to refer to solid-state spectroscopic studies of elementary excitations such as magnons and orbital excitations \cite{Ament2011}. In this article, we use RXES to denote core-level spectroscopy, and RIXS to refer to spectroscopy focused on the behavior of valence electron systems near the Fermi level (see Fig.~\ref{fig:schematic}).

One of the most important developments in RIXS was the recent experimental and theoretical realization of its sensitivity to spin-flip scattering processes,\cite{deGroot1998,Kuiper1998,Chiuzbaian2005,Hill2008,Ellis2010} including propagating magnon excitations \cite{Braicovich2009,Ament2009,Haverkort2010}.
There have been a large number of RIXS studies of various cuprate compounds carried out at the copper $L_3$ edge \cite{Schlappa2009,LeTacon2011,Schlappa2012,Dean2012,Dean2013a,Dean2013b,Dean2015}. Partially motivated by these activities, similar $L_3$ edge RIXS experiments were carried out to study $\rm Sr_2IrO_4$, in which crystal field excitations between the $t_{2g}$ and $e_g$ electronic states were observed to have a large resonance enhancement \cite{Ishii2011}. However, due to poor energy resolution, it was not possible to resolve low energy electronic or magnetic excitations in this early investigation. Thanks to the improved instrumentation, the first RIXS measurements of magnetic excitations in iridates were carried out at the Advanced Photon Source in 2010 \cite{JKim2012a}, which showed spectacular dispersion of both magnons and a spin-orbit exciton mode, the latter arising from the strong spin-orbit coupled magnetic degree of freedom. We note that iridium has a large neutron absorption cross-section, and studying iridates with inelastic neutron scattering is difficult. Most of the iridate neutron scattering work so far have been limited to elastic scattering.\cite{Dhital2013,Ye2012,Choi2012}
As a result, collective magnetic excitations in iridates have been largely elucidated by RIXS experiments. In addition to $\rm Sr_2IrO_4$ \cite{JKim2012a}, there have been a number of RIXS investigations of magnetic excitations in undoped and doped $\rm Sr_3Ir_2O_7$ \cite{JKim2012b,MorettiSala2015,Hogan2016,Lu2017}, thin film $\rm Sr_2IrO_4$ \cite{Lupascu2014}, doped $\rm Sr_2IrO_4$ \cite{Liu2016,Gretarsson2016,Pincini2017,Cao2017}, superlattice of $\rm SrIrO_3$ and $\rm SrTiO_3$ \cite{Meyers2017},
$\rm Na_2IrO_3$\cite{Gretarsson2013b}, $\rm Sm_2Ir_2O_7$ \cite{Donnerer2016}, and $\rm Eu_2Ir_2O_7$ \cite{Clancy2016,Chun2018}.

In addition to collective magnetic excitations, RIXS is an excellent probe of broadly-defined $d-d$-excitations. These are transitions between different d-orbitals, and range from highly dispersive collective excitations called orbitons to localized atomic transitions. If one only considers dipole transitions in the atomic limit, the direct $d$ to $d$ transition is forbidden optically. However, the $d$-$d$ transition is allowed and strong in the case of RIXS. This is because in RIXS, the $d$ to $d$ transition occurs through a second order process involving a dipole allowed $p$ to $d$ transition followed by a $d$ to $p$ transition. Detailed characteristics of these $d-d$ excitations differ from one material to another, depending on the various energy scales, such as electron bandwidth, electron correlation, crystal field splitting, and the strength of spin-orbit coupling.

In many iridates, the octahedral crystal field is usually large enough to keep Ir$^{4+}$ ($5d^5$) ions in the low-spin $t_{2g}^5$ state, allowing one to use $d-p$ isomorphism and map three t$_{2g}$ states to an effective $l_{eff}=-1$ state.\cite{Abragam1951} The spin-orbit coupling with strength $\lambda$ acting on the $t_{2g}^5$ manifold splits the 6 $t_{2g}$ states into a half-filled $j_{eff}=1/2$ doublet and a fully-filled $j_{eff}=3/2$ quartet; the energy difference between these two levels is $3 \lambda /2$.\cite{Hozoi2014} Therefore, the nature of $d$ to $d$ excitations changes from a simple orbital transition to a transition between spin-orbit coupled $j_{eff}$ states as $\lambda$ increases. Of course, this simple atomic picture is modified in real materials, with additional distortion of the octahedra giving rise to non-cubic crystal field and further splitting the $j_{eff}$ states. In particular, when there is a significant hopping between Ir, the transitions between $j_{eff}$ states become a collective excitation called a spin-orbit exciton.\cite{JKim2012a} Since the exchange interaction responsible for the propagation of the spin-orbit exciton has the same origin as the magnetic exchange coupling (i.e., hopping amplitude), one observes a large dispersion of the $d-d$-excitation only when magnetic exchange coupling is large. The $d-d$-excitations in iridates have been extensively studied \cite{Liu2012,Gretarsson2013a,Hozoi2014,MorettiSala2014a}. In a recent RIXS study of $\rm Sr_2IrO_4$, Kim et al. were able to extract separate dispersion for different spin-orbit exciton modes utilizing the different photon polarization dependence of these modes \cite{JKim2014}.

On the theory side, an important recent development has been the advances in {\em Ab Initio} quantum chemistry calculation methods to accurately determine the $d-d$ excitation energy. This method was applied to $\rm Na_2IrO_3$, $\rm Sr_3IrCuO_6$, and $\rm CaIrO_3$, which showed that $j_{eff}=1/2$ description works well in the former case\cite{Gretarsson2013a}, but not so well in the latter two materials \cite{Liu2012,MorettiSala2014a}. In the case of pyrochlore iridates, such as $\rm Eu_2Ir_2O_7$ and $\rm Y_2Ir_2O_7$, combined RIXS and quantum chemistry study found that non-cubic contributions to the crystal field splitting are quite significant in all pyrochlore materials studied, and the $j_{eff}$=1/2 description is somewhat questionable \cite{Hozoi2014}. In addition, Hozoi et al. found that the trigonal distortion of the $\rm IrO_6$ octahedra is not the major contributor to this non-cubic crystal field splitting in pyrochlore iridates. Rather it was found that the next nearest neighbor Ir and Eu/Y ions provide symmetry breaking field that causes the $t_{2g}$ levels to split. In their recent study of layered $\rm Sr_2IrO_4$, it was also found that the next nearest neighbors contribute significantly in determining crystal field levels in this material\cite{Bogdanov2014}.

On the other hand, $d-d$ excitations in iridates with oxidation states other than Ir$^{4+}$ have not been investigated much.\cite{Yuan2017} Materials with Ir$^{5+}$ and Ir$^{3+}$ oxidation state with $5d^4$ and $5d^6$ electronic configurations, respectively, are expected to show $d-d$ excitation spectra quite different from that observed for the $5d^5$ systems mentioned above. An additional energy scale, Hund's coupling ($J_H$), needs to be considered in the case of the $5d^4$ system, and in some cases, RIXS spectra can be analyzed to estimate both $J_H$ and $\lambda$ with high precision.\cite{Yuan2017} We also expect that the study of electronic states in $d^4$ and $d^6$ configurations will shed light on the hole- or electron-doping of the $d^5$ state, which is drawing much attention recently.\cite{Liu2016,Gretarsson2016,Pincini2017,Cao2017} Since electronic states near the Fermi level are dominated by Ir $d$ states, hole or electron-doping of the $d^5$ state would be equivalent to introducing $d^4$ (Ir$^{5+}$) or $d^6$ (Ir$^{3+}$) states, respectively, in the ionic picture.

Despite intense interest and many RIXS investigations of iridates, a systematic study of the RIXS cross-section in various iridates, including incident energy dependence, is still lacking. Studying Ei dependence of RIXS is useful for understanding the resonant process, as shown in the case of cuprates and other transition metal compounds.\cite{Lu2006,Ghiringhelli2009,Harada2000}
There have been only a limited number of incident energy dependence studies, for example, in $\rm Sr_2IrO_4$ \cite{Ishii2011}, $\rm Na_2IrO_3$ \cite{Gretarsson2013b}, and $\rm CaIrO_3$ \cite{MorettiSala2014a}.

Even fewer core-level RXES studies have been carried out for iridates except for $\rm CuIr_2S_4$ \cite{Gretarsson2011}. RXES can give us detailed information about the valence of Ir atoms. This is particularly important for studying mixed valence iridate compounds such as $\rm Ca_5Ir_3O_{12}$. The usual experimental method for studying oxidation state, x-ray absorption spectroscopy (XAS) is not very informative in the case of Ir $L_3$ edge, since core-hole lifetime effect broadens spectral features. The RXES and XAS measured using high energy-resolution fluorescence detection (HERFD) can overcome this limitation, and both of these methods are employed in the present study.

The purpose of the current work is twofold. First, we would like to focus on higher energy excitations for these compounds often ignored in previous RIXS studies, and in particular, provide a systematic view of the $d-d$-excitation spectra of iridates by comparing various iridate compounds. The second purpose of the current work is to focus more on the spectroscopic aspect of iridate RIXS. Instead of focusing on momentum dependence, which is important for understanding solid-state phenomena, here we focus more on the incident energy dependence and the x-ray absorption and emission processes, which are best understood in the ionic/atomic picture. Of course, we also acknowledge that even these $d-d$ excitations have significant contributions from neighboring atoms and even further lying ions, as recently pointed out in Ref.~\onlinecite{Bogdanov2014}, However, our interest here is a phenomenological description and for the most part,  we will rely heavily on local atomic descriptions.

In the next section, we will first discuss experimental details of our spectrometer setup, and also the samples used in our study. Experimental results and discussions will be divided into two sections. First, we will present RXES and HERFD results. In the following section, RIXS data will be presented. This section will be further divided into two subsections discussing  $E_i$-dependence and $d-d$ excitations. The low energy $d-d$-excitations below 1~eV have been the subject of several earlier publications, and therefore will not be discussed in detail. Due to the broad nature of the topics presented in this article, results and discussions are presented together in each subsection.

\section{Experimental methods}

\subsection{Instrumentation}
The RXES experiments were carried out at the Advanced
Photon Source using 9ID RIXS spectrometer, utilizing the same setup used in the previous RXES study of $\rm CuIr_2S_4$.\cite{Gretarsson2011}
The beam
was monochromatized by a double-crystal Si(111) and a
Si(333) channel-cut secondary crystal to select incident photon energy near the Ir $L_3$ edges. A spherical (1-m-radius) diced Ge(337)
analyzer was used to select final photon energy near the Ir L$\alpha_2$ ($3d \rightarrow 2p$) emission line at 9.1 keV. The scattering plane was vertical. The overall energy resolution [full width at half maximum
(FWHM)] in this configuration was about 240 meV.
The XAS data using high energy-resolution fluorescence detection method
(HERFD-XAS) was also collected using the same setup.

Two different experimental setups were used for the RIXS experiments at the 9ID beamline.
Most of the incident energy dependence and the dd excitation spectra presented in Sec.~\ref{sec:rixs} were obtained using the lower-resolution setup, which was similar to the RXES setup described above, but with a Si(844) channel-cut secondary monochromator and a spherical (1-m-radius) diced Si(844) analyzer. In order to
minimize the elastic background intensity, measurements
were carried out in the horizontal scattering geometry, for which the scattering angle $2 \theta$ was close to
90 degrees. The overall energy resolution of about 150 meV (FWHM)
was obtained. Low-lying $d-d$-excitations below 1~eV were measured using a higher resolution setup. By using a better-focused x-ray beam, an overall energy resolution of
80 meV (FWHM) and a significant reduction of the elastic
line could be achieved.

\subsection{Samples}

\begin{table*}
\caption{\label{table:samples} List of samples studied in this work. We list a brief description of the structure as well as whether we studied powder or single crystal samples. We also list the nearest-neighbor Ir-Ir bond length for each crystal structure. The Ir-Ir bond lengths in the Rh-doped samples are similar to those of undoped $\rm Sr_2IrO_4$. If the sample was used in a previous  publication, the reference is listed here.}
\begin{ruledtabular}
\begin{tabular}{cccccc}
&Structural Motif&Ir-Ir(\AA)&Form&Reference\footnote{Previous RIXS study, if any.}
&Note\\ \hline
 Ir&fcc&2.71&powder&  &commercial \\
 $\rm IrCl_3$&honeycomb&3.45&xtal& & Ir$^{3+}$\\
 $\rm IrO_2$&rutile&3.16&powder& & commercial\\
 $\rm Li_2IrO_3$& honeycomb &2.98& powder& \onlinecite{Gretarsson2013a} & \\
 $\rm Sr_2IrO_4$& square &3.89& xtal& \onlinecite{JKim2012a} & \\
 $\rm Sr_2Ir_{0.89}Rh_{0.11}O_4$& square && xtal& \onlinecite{Clancy2014} & \\
 $\rm Sr_2Ir_{0.58}Rh_{0.42}O_4$& square && xtal& \onlinecite{Clancy2014} & \\
 $\rm Sr_2Ir_{0.3}Rh_{0.7}O_4$& square && xtal& \onlinecite{Clancy2014} & \\
 $\rm Ca_5Ir_3O_{12}$& chain &3.19&xtal & &mixed valence\\
 $\rm Eu_2Ir_2O_7$& pyrochlore &3.63& powder& \onlinecite{Hozoi2014} & \\
$\rm Sr_2YIrO_6$& double perovskite &5.79& xtal& \onlinecite{Yuan2017} & Ir$^{5+}$\\
\end{tabular}
\end{ruledtabular}
\end{table*}

In Table~\ref{table:samples}, samples studied in this investigation are listed and relevant information for each sample is provided. We used both single crystal samples and powder samples. We found that most of the $d-d$ excitations studied here (with the exception of $\rm Sr_2IrO_4$ based materials) have very little momentum dependence. This is due to the small magnetic exchange energy (order of 10 meV or less), which is too small to give a sizable dispersion for $d-d$ excitations compared to the instrumental resolution, thereby justifying our use of polycrystalline powder samples. For example, earlier studies showed that the RIXS spectra of single crystal Na$_2$IrO$_3$ and $\rm Eu_2 Ir_2 O_7$ are very similar to those measured using powder samples.\cite{Gretarsson2013a}
Powder samples of $\rm Li_2 Ir O_3$ and $\rm Eu_2 Ir_2 O_7$ have been synthesized using the standard solid-state reaction method.
The IrO$_2$ and Ir metal samples are commercial samples purchased from Alfa Aesar (99.99 \%) and used without further treatment.

The Ir$^{3+}$ reference sample, $\alpha$-IrCl$_3$, crystallizes in monoclinic structure C2/m \cite{Brodersen1965}. Single crystal samples were synthesized using vacuum sublimation as described in Ref.~\onlinecite{Brodersen1965}. Small dark yellow transparent crystals are used in our measurement. IrCl$_6$ octahedra share edges to form a layer of honeycomb net, which is stacked in monoclinic fashion. This structure is rather similar to $\alpha$-RuCl$_3$, which is drawing much interest due to the magnetic behavior associated with Kitaev-Heisenberg Hamiltonian \cite{Plumb2014,Sears2015,Majumder2015,Sandilands2015,Kubota2015,HSKim2015,Banerjee2016}. Unlike its ruthenium cousin, the Ir$^{3+}$ has filled t$_{2g}$ states ($d^6$), which makes this material a band insulator with a large insulating gap. 
Due to the plate-like crystal morphology, the experiment was carried out with photon polarization within the hexagonal plane.

The same single crystal sample of ordered double perovskite $\rm Sr_2YIrO_6$ used in the earlier RIXS investigation was used here.\cite{Yuan2017} Details of the crystal growth was reported in Ref.~\onlinecite{GCao2014}. We note that in this sample the iridium ions have Ir$^{5+}$ ($5d^4$) valence. Since non-magnetic Y$^{3+}$ is placed between two Ir ions, the nearest-neighbor Ir-Ir distance in this material is much larger than that in other materials. The nearest-neighbor Ir-Ir bond lengths are compared in Table~\ref{table:samples}. Except for Ir and $\rm Sr_2YIrO_6$, Ir is connected to its neighbor Ir through O (or Cl) ligand ions, and the variation in the bond lengths just reflect the difference in the Ir-O-Ir angles in these compounds. Due to the long-range nature of hopping, we expect the hopping integral to be much suppressed in $\rm Sr_2YIrO_6$.

$\rm Sr_2IrO_4$ has been studied with RIXS quite extensively, and a comprehensive doping dependence of the low energy magnetic excitation in doped $\rm (Sr,La)_2(Ir,Ru)O_4$ is available elsewhere \cite{Liu2016,Gretarsson2016,Pincini2017,Cao2017}. The samples studied in this paper are the same samples used in earlier magnetic diffraction and transport experiments \cite{Qi2012,Clancy2014}. We will present here the incident energy dependence and high energy charge excitations which have not been reported in earlier investigations. We note that the incident energy study of the undoped sample using coarse energy resolution was reported in Ref.\ \onlinecite{Ishii2011}.

We also investigated a mixed valence compound \CIO. This compound was initially identified as $\rm Ca_2 IrO_4$ phase \cite{Sarkozy1974}, but later structural refinements found that the correct stoichiometry is \CIO and the structure is described as hexagonal with $P\bar{6}2m$ symmetry \cite{Dijksma1993,Cao2007}. Single crystals were synthesized using the method reported in Ref.~\onlinecite{Cao2007}. This material has a chain motif, and magnetic order was observed below about 7~K. Since the average valence of Ir in this compound is Ir$^{4.67+}$, one would naively expect mixed valence of about 2/3 Ir$^{5+}$ and 1/3 Ir$^{4+}$. We do observe a mixture of 4+ and 5+ valence states in this compound.

\section{Results and discussion: RXES}
\label{sec:rxes}

\begin{figure*}
\begin{center}
\includegraphics[angle=0,width=7in]{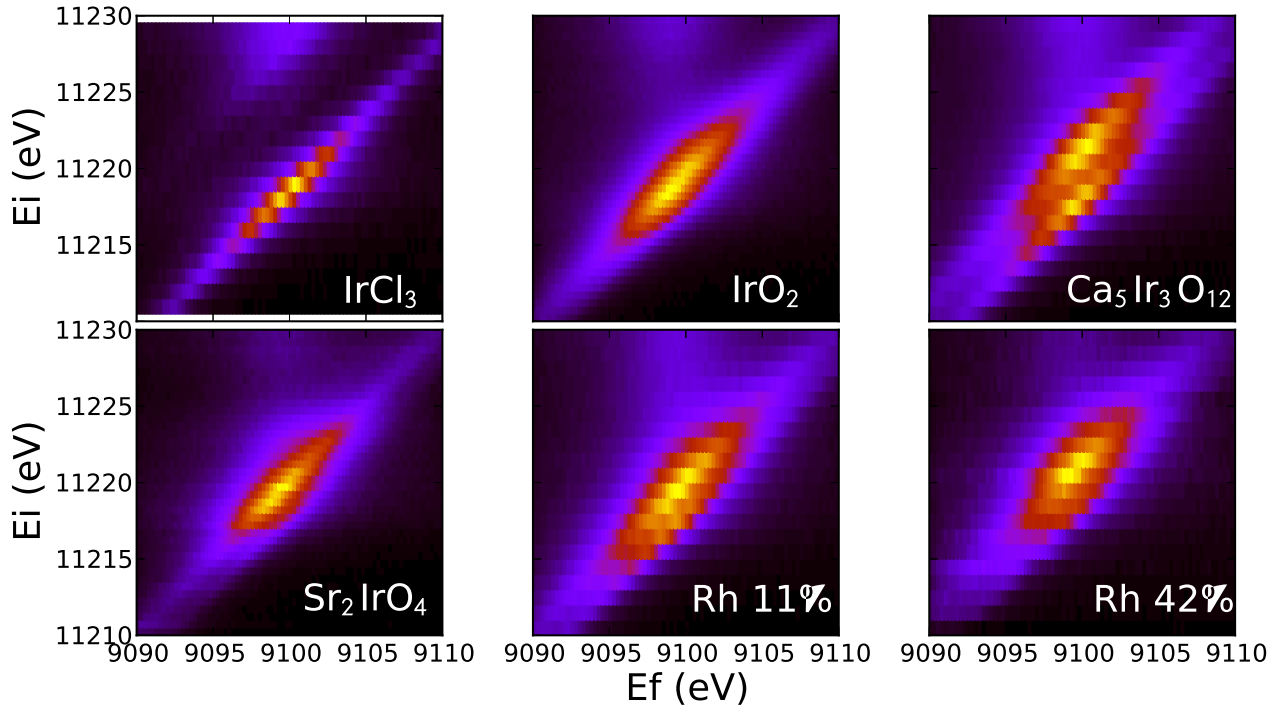}
\end{center}
\caption{(Color online) Resonant X-ray Emission Spectroscopy (RXES) map for Ir L$\alpha 2$ emission in various iridate compounds. The RXES intensity is plotted as a function of both incident energy and detected energy for various iridate compounds. A false color scale is used in this plot. Light/bright color means higher intensity.}
\label{fig:rxes}
\end{figure*}

In Fig.~\ref{fig:rxes}, the intensity of the Ir L$\alpha_2$ emission for various iridate compounds is shown as a two-dimensional map of $E_i$ and $E_f$. \cite{deGroot_book} There are two major spectral features observed in this intensity map. One is an intense resonant feature with characteristic elliptical shape, and the other is a faint streak of intensity at higher $E_i$ values. These two features exist in all six samples, but perhaps it is easiest to see this in the case of IrO$_2$. The difference between these two features is their dependence on $E_f$. The strong feature occurs when the transferred energy ($E_i-E_f$) takes on a definite value, similar to the Raman shift observed in a light scattering experiment. That is, as the incident energy increases, the emission occurs at correspondingly higher energy. This results in the elliptical shape of the excitation in this $E_i$ vs. $E_f$ map, in which the slope of the major axis of the ellipse is one (i.e., 1 eV change in $E_i$ results in 1 eV change in $E_f$). On the other hand, the weak higher $E_i$ feature occurs at a constant $E_f$ value, and this corresponds to normal fluorescence. This type of two-feature behavior is commonly observed in RXES studies.\cite{Rueff2010,Doring2004} In our study, we focus only on the strongly Raman-like feature with an elliptical shape.

These RXES maps are very useful for elucidating the oxidation state of ions. One of the best known methods to study oxidation state is x-ray absorption spectroscopy (XAS), which is traditionally obtained by measuring transmission through thin samples or detecting fluorescence yield from the sample as incident photon energy is varied across the absorption edge. In the latter case, a typical solid-state detector with an energy resolution of a few hundred eV is used to detect the intensity of a particular fluorescence line, or several lines together, which is often called a fluorescence yield (FY) mode. However, both methods suffer when the absorption line width becomes extremely broad due to very short core-hole lifetime. This is the case in iridate materials; for example, the width of 2p$_{\frac{3}{2}}$ core level in Ir is 5.25 eV, which is much broader than that of 3d transition metals like Cu (0.56~eV).\cite{Krause1979} As a result, the XAS spectra at Ir L-edges are very broad without clear spectral features to speak of.\cite{Clancy2012} There have been several efforts to address this shortcoming of the XAS method. In their pioneering work, Hamalainen and coworkers were able to obtain much sharper XAS spectra of Dysprosium by utilizing a high resolution analyzer to detect only emitted photons with a specific energy.\cite{Hamalainen1991} Since then there have been numerous theoretical and experimental studies exploiting this HERFD-XAS technique.\cite{deGroot_book}

\begin{figure}[!ht]
\begin{center}
\includegraphics[angle=0,width=3.1in]{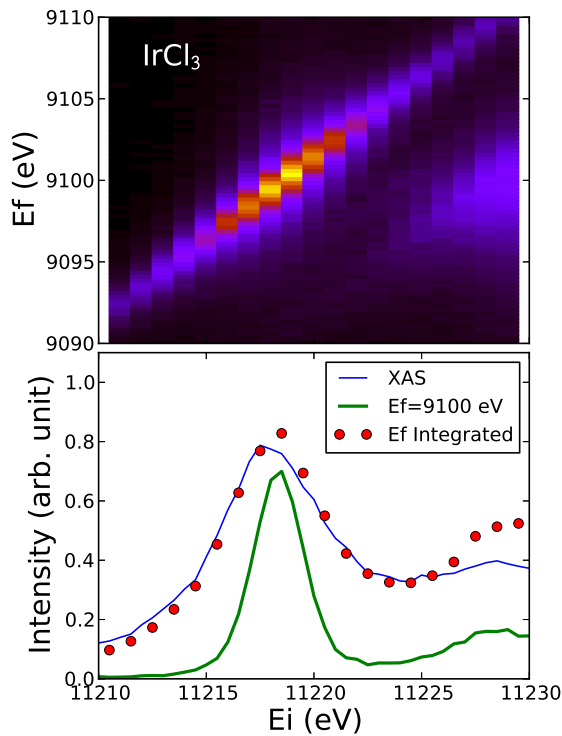}
\end{center}
\caption{(Color online) X-ray absorption near edge spectra for IrCl$_3$. The RXES map shown in Fig. 2 is transposed here to show $E_i$ in the horizontal axis. The thin blue line is the XAS data obtained with a solid-state detector (Amptek). The data points for fixed analyzer energy $E_f=9100$~eV are plotted in the thick green line. The circles correspond to the intensity for given $E_i$ integrated over all $E_f$ from 9090~eV to 9110~eV).}\label{fig:ircl3}
\end{figure}

\begin{figure}[!ht]
\begin{center}
\includegraphics[angle=0,width=3.1in]{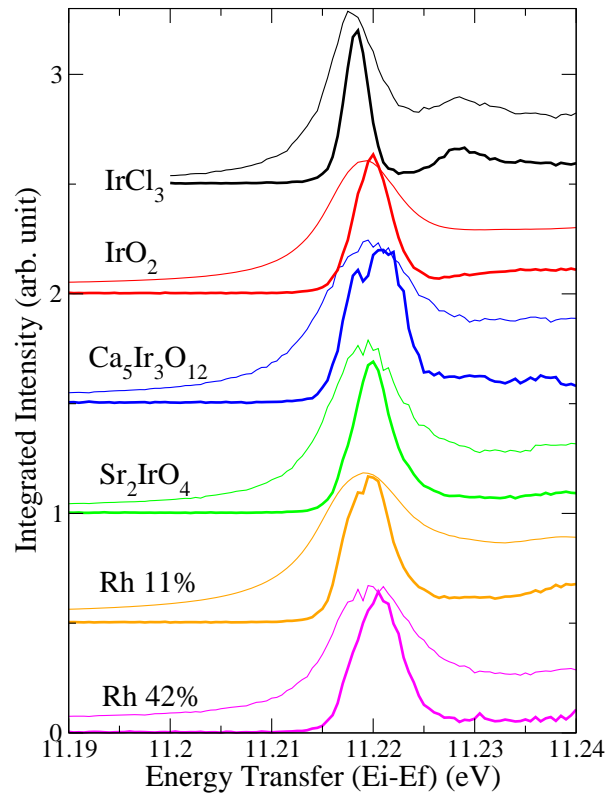}
\end{center}
\caption{(Color online) X-ray absorption near edge spectra for various iridates are compared. The spectra obtained using the FY and HERFD mode are shown in thin and thick lines, respectively. The HERFD mode data are the Ei dependences of the intensity when the analyzer was fixed to $E_f=9100$~eV as illustrated for IrCl$_3$ in Fig.~3.}\label{fig:pfy}
\end{figure}

The reason why one can reduce the core-hole lifetime broadening using the high resolution analyzer method can be easily understood by inspecting the RXES map. In Fig.~\ref{fig:ircl3}, we illustrate this using the example of IrCl$_3$. The horizontal and vertical axes are flipped in this case to show $E_f$ vs. $E_i$ map. A typical XAS experiment using a detector with a coarse energy resolution is equivalent to having no capability of distinguishing $E_f$ values in this figure. To simulate regular XAS spectra we integrated over all $E_f$ values and plotted this in Fig.~\ref{fig:ircl3} as filled circles. Indeed, this result agrees very well with the XAS data obtained using a solid state detector shown as a solid line in Fig.~\ref{fig:ircl3}(b).
However, a high resolution analyzer enables one to select a particular $E_f$ value, and by doing so, one can reduce the apparent linewidth of the XAS spectra as shown here. If we plot only the intensity corresponding to the $E_f$=9100~eV, we obtain the thick green solid line shown in Fig.~\ref{fig:ircl3}(b). While the observed width of the integrated intensity (or the FY-XAS width) is roughly 5 eV as expected, this is reduced by about a factor of 2 in the HERFD-XAS spectrum obtained by selecting $E_f$=9100~eV. This observed width of 2.7 eV is much sharper than the Ir 2p lifetime (5.25 eV) and close to the $3d_{3/2}$ lifetime of about 2 eV as described below.
In Fig.~\ref{fig:pfy}, we compare the XAS data for various iridate samples obtained using the HERFD method with those obtained with the normal fluorescence method.

While the normal XAS results are more or less indistinguishable from one another, the HERFD-XAS results show some interesting structures in the main absorption peak (the so-called white line). The first thing to note is the clear shift of the peak position in the case of IrCl$_3$, which is the only compound with +3 valence. This type of edge shift in XAS is commonly observed, and used to distinguish oxidation states in mixed valence compounds. In the case of $3d$ transition metal compounds, van der Laan and Kirkman showed that the shift mostly originates from the difference in the Coulomb interaction between the $2p$ core hole and $3d$ valence electrons. \cite{Laan1992} Small edge shifts of order $\sim 1$~eV were observed going from Ir$^{4+}$ to Ir$^{6+}$ in the conventional XAS data reported by Laguna-Marco et al. on a series of Ir-based double perovskites.\cite{LagunaMarco2015} Instead of a shift, what is clearly observed in our HERFD-XAS data on \CIO and Rh-doped $\rm Sr_2IrO_4$ is the broadening of the absorption feature when Ir$^{5+}$ is introduced. Note that \CIO has nominally 67 \% Ir$^{5+}$, and some Ir$^{5+}$ are also present in the Rh doped samples. If one compares these compounds to IrO$_2$ and $\rm Sr_2IrO_4$, both with purely +4 valence, there seems to be an additional feature on the lower energy side. For \CIO, there is a clear splitting between two peaks, while Rh doped $\rm Sr_2IrO_4$ shows broadening. Although it is tempting to attribute the splitting to +4 and +5 valence, things are a little more complicated due to the large crystal field splitting of about 3 eV between the t$_{2g}$ and e$_g$ states as described below.

The low-energy feature (split peak or broadening due to a shoulder) originates from the increased transition probability into Ir $t_{2g}$ states in Ir$^{5+}$ ions. At the Ir $L_3$ edge, 2p$_{3/2}$ core electrons are excited into empty 5d states. For IrCl$_3$, the only available states are e$_g$ states, and the line width probably represents the e$_g$ band width in IrCl$_3$. For Ir$^{4+}$ ions, nominally one empty $t_{2g}$ state becomes available (compared to 4 $e_g$ states per ion), and the peak becomes broader. The $\rm Sr_2IrO_4$ data indeed looks like there is a weak shoulder feature at lower energy side. As you increase Ir$^{5+}$ composition, the number of empty $t_{2g}$ states increases and the shoulder feature becomes more prominent. Note that at $L_3$ edge, there is no selection rule preventing transitions involving spin-orbit coupled $t_{2g}$ states unlike at the $L_2$ edge.

\begin{figure}
\begin{center}
\includegraphics[angle=0,width=3.1in]{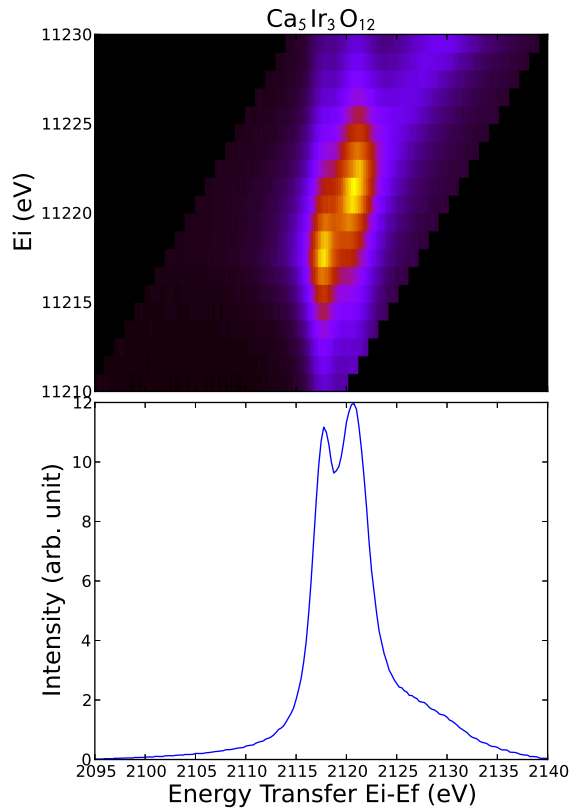}
\end{center}
\caption{(Color online) (a) Intensity plotted as a function of incident photon energy and energy transfer for \CIO, the same data shown in Fig.~2. (b) The intensity integrated over $E_i$ plotted as a function of energy transfer.}\label{fig:cio}
\end{figure}

\begin{figure}
\begin{center}
\includegraphics[angle=0,width=3.1in]{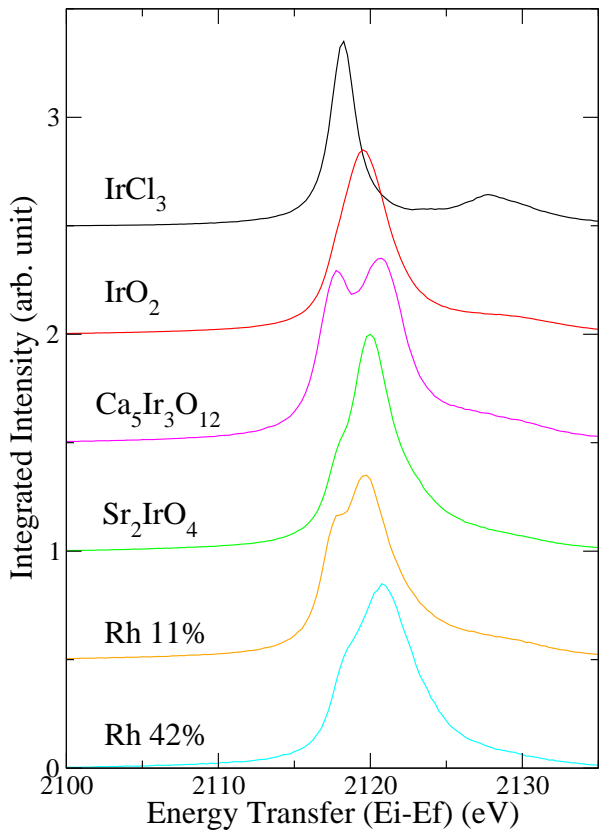}
\end{center}
\caption{(Color online) (a) ``Quasi $M_4$" absorption edge spectra obtained by RXES method described in Fig.~5 are shown for various iridates studied.}\label{fig:m4}
\end{figure}

An alternative way to visualize the RXES data is to plot $E_i$ against the energy transferred to the system ($E_i-E_f$) as illustrated for \CIO in Fig.~\ref{fig:cio}. This intensity map clearly shows that the resonant spectral features occur at constant energy transfer. One should note that the transitions involved in this experiment are $2p_{3/2} \rightarrow 5d$ followed by $3d_{3/2} \rightarrow 2p_{3/2}$. The final state of this RXES process, therefore, is identical to removing an electron from the $3d_{3/2}$ state, which would be similar to $M_4$ edge XAS.\footnote{True $M_4$ XAS final state would have a hole in $3d_{3/2}$ and an extra electron in $6p-5f$ states instead of $5d$ states.} In this case integrating over $E_i$ is equivalent to integration over all final states, since different $E_i$ corresponds to different excited (final) $5d$ states of the $M_4$ absorption process. We can obtain ``quasi'' XAS spectra for the Ir $M_4$ edge as shown in the bottom panel by integrating over all $E_i$ values. This ``quasi $M_4$-edge'' XAS shows even clearer splitting between the two peaks corresponding to transitions to t$_{2g}$ and e$_g$ levels.
In Fig.~\ref{fig:m4}, such quasi $M_4$-edge XAS spectra are shown for the samples considered in this study. The information contained in this plot is equivalent to the one in Fig.~\ref{fig:pfy}, but the linewidths are much narrower, reflecting the lifetime of 3d electronic states. We fitted the results shown in Fig.~\ref{fig:m4} to a two-peak Lorentzian lineshape. The fitting results are listed in Table~\ref{table:M4}. The overall trend is quite similar to the HERFD-XAS data, but sharper spectral features allow one to make quantitative comparisons with future calculations. One thing we would like to comment on is the behavior of Rh doped samples seem to be somewhat anomalous, since the peaks in the Rh-11\% sample are at slightly lower energy than either undoped or Rh-42\% samples. In their recent investigation of the doping dependence of x-ray absorption spectra (conventional FY method), Chikara and coworkers found that the evolution of charge partitioning between Rh and Ir with Rh doping is anomalous. Specifically, Rh doping introduces holes in the Ir sites at low doping, but at high doping Rh takes up the holes and the fraction of Ir$^{5+}$ ions stay below 25\%. Our results may reflect this anomalous behavior, but further systematic studies are clearly needed to gain quantitative information about this.

\begin{table}
\caption{\label{table:M4}
Fitting results for the peak positions and widths from the spectra shown in Fig.~6. The Lorentzian position and the full width of the absorption feature associated with the empty t$_{2g}$ and e$_g$ states are listed. Only a single peak fitting was done for IrO$_2$, since no clear peak splitting is observed.}
\begin{ruledtabular}
\begin{tabular}{ccccc}
&t$_{2g}$ (eV)& width (eV)&e$_g$ (eV) & width (eV)\\ \hline
 $\rm IrCl_3$& - & - & 2118.2 & 2.0 \\
 $\rm IrO_2$& - & - & 2119.6 & 3.8  \\
 $\rm Ca_{5}Ir_3O_{12}$& 2117.7 & 2.2 & 2120.7 & 4.0 \\
$\rm Sr_2IrO_4$& 2118.4  & 2 & 2120.3 & 3.4\\
Rh 11 \% & 2117.8 & 2 & 2120.0 & 3.6\\
Rh 42 \% & 2118.7 & 2 & 2121.1 & 3.8\\
 \end{tabular}
\end{ruledtabular}
\end{table}

\section{Results and discussion: RIXS}
\label{sec:rixs}

\subsection{$E_i$ dependence}

\begin{figure}[ht!]
\begin{center}
\includegraphics[angle=0,width=3.2in]{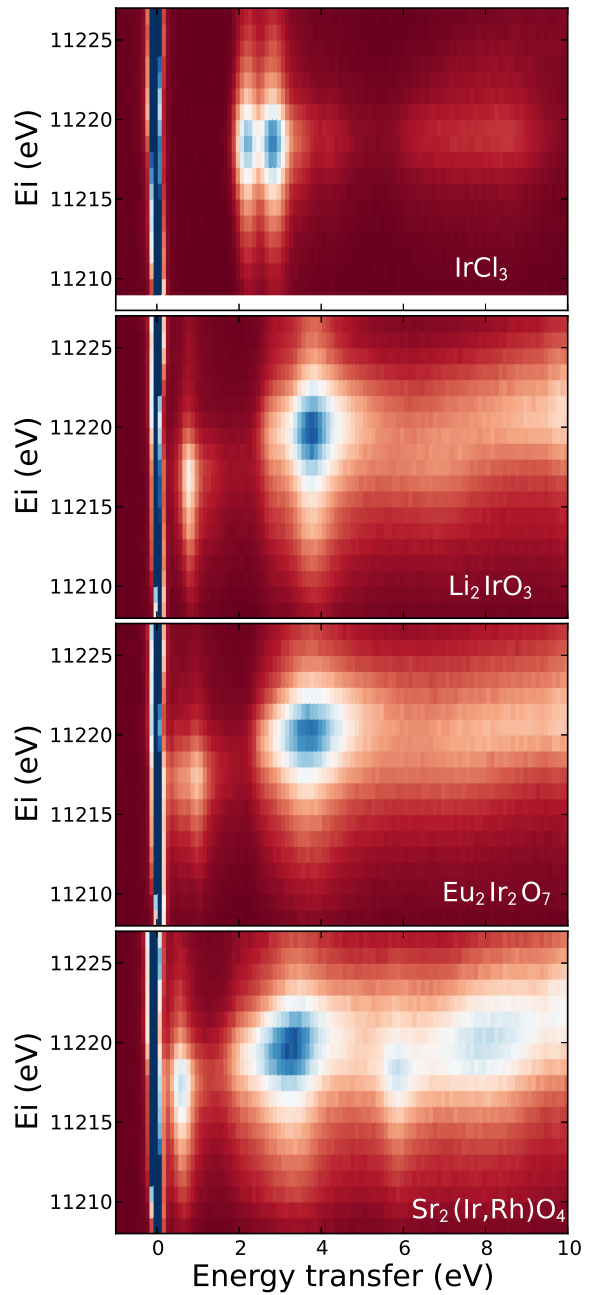}
\end{center}
\caption{(Color online) Resonant Inelastic X-ray Scattering (RIXS) map for valence excitations in various iridate compounds. The RIXS intensity is plotted as a function of both incident energy and energy transfer ($E_i - E_f$). A false color scale is used in this plot. Blue color means higher intensity, and red color means lower intensity.}\label{fig:rixs}
\end{figure}

In Fig.~\ref{fig:rixs}, the incident energy dependence of the RIXS spectra for various iridates are shown. The RIXS intensity is measured as a function of energy transfer, which corresponds to the excitation energy of electrons in the sample. The momentum transfers are fixed for these measurements. The incident energy dependence is also easy to visualize by plotting the RIXS intensity map as shown in this figure. The intensity scale is in an arbitrary unit, since the absolute intensity comparison between different samples is difficult.

\begin{figure}
\begin{center}
\includegraphics[angle=0,width=3.1in]{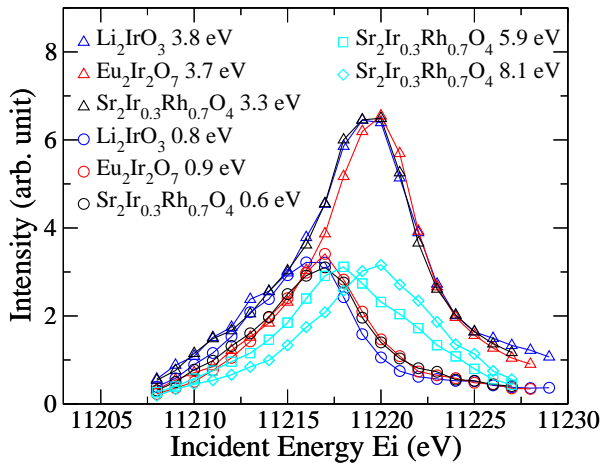}
\end{center}
\caption{(Color online) Incident energy dependence of the intensity of various excitations shown in Fig.~7. Different symbols are obtained from vertical cuts from Fig.~7 for fixed energy transfer.}\label{fig:res}
\end{figure}

RIXS is a second order process involving an intermediate state (state $|n \rangle$ in Fig.~\ref{fig:schematic}). RIXS intensity is usually calculated using the Kramers-Heisenberg formula:
\begin{equation}
I \sim \sum_j  \left| \sum_n \frac{ \langle j | T | n \rangle \langle n | T | g \rangle}{ E_g + E_i -  E_n - i \Gamma_n} \right| ^2 \delta(E_i+E_g-E_n-\hbar \omega),
\end{equation}
where $E_i-E_f=\hbar \omega$ is the energy transfer, and $\Gamma_n$ is the intermediate state lifetime. Since the intensity is enhanced when the incident energy is tuned near the intermediate state energy ($E_i \approx E_n-E_g$), inspection of the incident energy dependence of each excitation can help identify the particular intermediate state involved in the resonant enhancement. Since these intermediate states correspond to the final state of XAS, one can gain insight into the intermediate states by comparing the incident energy dependence and the XAS spectrum. In Sec.~\ref{sec:rxes}, we found that the XAS spectra are composed of empty t$_{2g}$ and e$_g$ state contributions; the empty t$_{2g}$ state (XAS final, RIXS intermediate) is found about 3 eV below the main peak in the absorption spectrum.

The most intense spectral feature in the RIXS map is around $\hbar \omega=3-4$~eV excitation energy, which resonates around $E_i=11220$~eV (the XAS peak positions in Fig.~4). This is the $d-d$ excitation between the t$_{2g}$ and e$_g$ states. For this excitation, the intermediate state would correspond to $\underline{2p}t_{2g}^5e_g^1$, where underline denotes having a core hole in the Ir $2p_{3/2}$ state; then one of the t$_{2g}$ electrons fall to fill the core hole, and the final state would be $t_{2g}^4e_g^1$ (see Fig.~\ref{fig:schematic}). As mentioned in the introduction, RIXS is particularly effective in detecting this type of $d-d$ excitation, which is made up of allowed dipole transitions. The other excitations between d-levels, in particular the excitation between non-degenerate t$_{2g}$ states occurs through an intermediate $\underline{2p}t_{2g}^6$ state, and therefore will occur at a lower $E_i$ value. As one can see in Fig.~\ref{fig:rixs}, these intra-t$_{2g}$ excitations can be observed in the energy region below $\hbar \omega=1$ eV, and as expected resonate at $E_i$ about 3~eV lower than the resonance energy of the e$_g$ excitation. This trend is quite universal in all the samples studied, except for IrCl$_3$, which is a $d^6$ system and does not have empty t$_{2g}$ states.

One should note that only these two states are involved in the iridate RIXS process. Even the high energy excitations observed above 6~eV show resonance behavior corresponding to the intermediate state similar to t$_{2g}$ or e$_g$ excitations. The origin of these higher energy excitations is unclear, but most likely due to the charge-transfer excitation from oxygen to iridium. We note that excitations in the similar energy range have been observed in the oxygen K-edge RIXS measurement of Sr$_2$IrO$_4$.\cite{Liu2015}
The Ir$^{3+}$ compound, IrCl$_3$, does not have an empty t$_{2g}$ state, and the intra-t$_{2g}$ transitions are absent. The existence of only two resonance energies is clearly illustrated in Fig.~\ref{fig:res}, in which the $E_i$ dependence of the main spectral features in select samples are shown. This plot is obtained by taking vertical cuts through chosen excitation energies in Fig.~\ref{fig:rixs}, and plotting the intensity as a function of $E_i$ (exact values of $\hbar \omega$ are noted in the figure). There are clearly only two resonance energies, around $E_i=11217$~eV and $E_i=11220$~eV. This is quite a striking observation, especially when compared to the resonance behavior (observed through $E_i$-$\hbar \omega$ map) of Cu K-edge RIXS spectra, which exhibit much more complex behavior.\cite{LCO,Lu2005} In Ref.~\onlinecite{Observations}, it was pointed out that the RIXS resonance behavior depends crucially on  the intrinsic width of the excitation when compared to the inverse core-hole lifetime. In most cuprate materials such as $\rm La_2CuO_4$, the intrinsic charge transfer excitation energy width is of the same order as the Cu 1s lifetime of 1-2 eV, and the excitation spectra observed for Cu K-edge RIXS exhibit relatively complicated resonance behavior. In some energy regions, one can observe the resonance $E_i$ varies continuously with excitation energy $\hbar \omega$. In contrast, Ir 2p lifetime ($\sim$5~eV) is much larger than the typical width of a $d-d$ excitation, and we observe resonance behavior very similar to the short core-hole lifetime limit shown in Fig.~4 of Ref.~\onlinecite{Observations}. The only exceptions are high energy excitations at $\hbar \omega$=5.9~eV and 8.1~eV in the case of $\rm Sr_2Ir_{0.3}Rh_{0.7}O_4$, which is a metallic sample. In this case, one expects drastic broadening of charge excitation widths due to mobile electrons. As a result, resonance behavior is no longer bimodal as found in other cases.

This observation brings up the issue of the response function of RIXS. In their study of RIXS cross-section, van den Brink and van Veenendaal showed that the matrix element of the RIXS cross-section can be factored out in Eq.~(1), leaving charge and spin response functions that only depend on the properties of the valence electron system.\cite{UCL-EPL} The caveat is that this approximation only holds in the limit of ultra-short core-hole lifetime, and it seems like iridates are great examples of materials where this approximation works well. Empirically, the $E_i$ dependence shown in Fig.~\ref{fig:rixs} supports this, since the choice of the incident photon energy changes only the intensity of the peak and does not affect the peak position.

\begin{figure}
\begin{center}
\includegraphics[angle=0,width=3.1in]{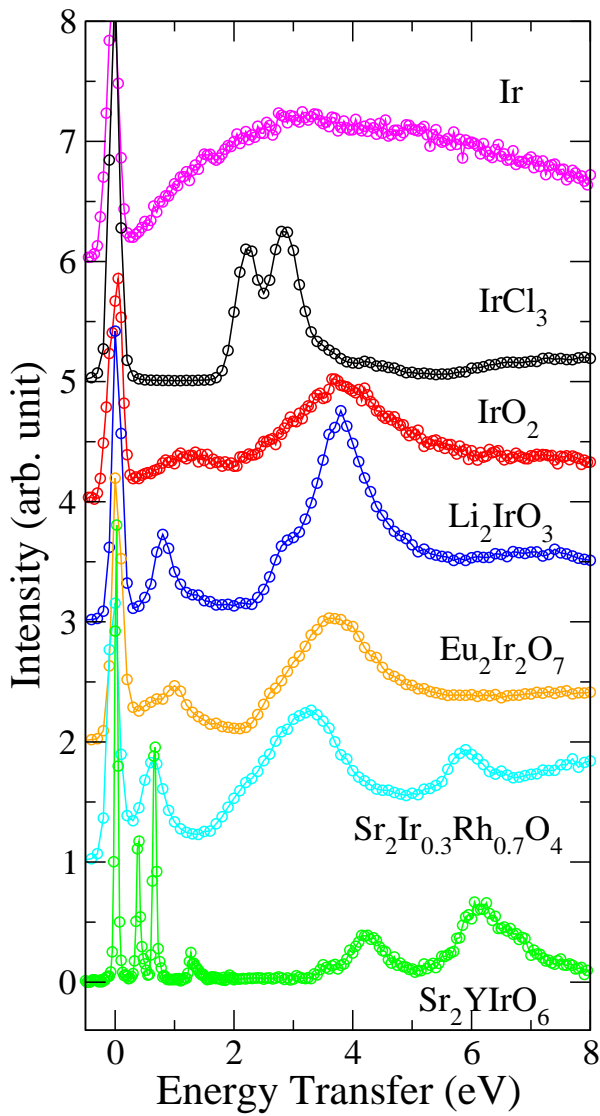}
\end{center}
\caption{(Color online) RIXS spectra showing dd excitations for various iridate samples. The incident energy was chosen to maximize the low energy excitation. $E_i=11215$~eV was used for $\rm Sr_2YIrO_6$, and $E_i=11219.5$~eV was used for Ir and IrO$_2$. $E_i=11218$~eV was used for all the other samples.}\label{fig:dd}
\end{figure}

\subsection{dd Excitations}

In Fig.~\ref{fig:dd}, RIXS spectra for  different iridate samples are compared. The incident energy was chosen to maximize the low energy intra-t$_{2g}$ excitations. Since the t$_{2g}$ to e$_g$ excitation has the rather strong intensity and broad resonance, it is also clearly visible even with the lower $E_i$. Thus we regard the spectra shown in Fig.~\ref{fig:dd} as rough representatives of electronic excitations in iridate samples over a wide energy range. Only very low energy magnetic excitations cannot be resolved from the elastic line.

What is noteworthy is the difference between Ir and all the other samples. Elemental Ir, perhaps not surprisingly, does not show any spectral features other than extremely broad intensity distribution over the whole energy range probed. This type of broad excitation spectra is commonly observed in metals and includes contributions from fluorescence.\cite{Magnuson2003} The IrO$_2$ spectrum also seems quite distinct from the other spectra. There are two broad features arising from the two types of $d-d$ excitations, reflecting the octahedral environment of the Ir atom in this sample.
All the other samples show qualitatively similar excitation spectra made up of low energy intra-t$_{2g}$ excitation, t$_{2g}$-e$_g$ $d-d$ excitation in the range of 2-5eV. There are some excitations at even higher energy above 6~eV, which may be due to multi electron transitions and/or charge transfer excitations, but we will not discuss these excitations in this paper. To aid our discussion we show a schematic representation of the density of states for an Ir$^{4+}$ system in Fig.~\ref{fig:schematic}(c).

All intra-t$_{2g}$ excitations more or less fall within the energy range 0.5-1 eV. This is expected since the primary energy scale determining the t$_{2g}$ splitting in iridates is the spin-orbit coupling ($\lambda$), which does not depend much on structural details. In contrast, the excitations between the occupied t$_{2g}$ states and empty e$_g$ states are quite broad, reflecting the wide bandwidths due to the strong hybridization between Ir e$_g$ and oxygen 2p states. In addition to the widths, we note that the positions of the t$_{2g}$-e$_g$ excitations are more or less unchanged across different materials, except for $\rm IrCl_3$ and $\rm Sr_2YIrO_6$, for which Ir has +3 and +5 nominal oxidation states, respectively. A simple electrostatic consideration can qualitatively account for this difference. Larger crystal field splitting for Ir$^{5+}$ than for Ir$^{4+}$ is expected since cation-anion electrostatic attraction will be stronger for more positive Ir$^{5+}$ ions. This also results in shorter Ir-O bond lengths for $\rm Sr_2YIrO_6$ (1.97 \AA\ for $\rm Sr_2YIrO_6$ and 2.02 \AA\ for $\rm Na_2IrO_3$). On the other hand, $\rm IrCl_3$ is expected to have weaker electrostatic interaction due to the net charge, and also due to the larger ionic size of chlorine ions. Not surprisingly the Ir-Cl bond length is much larger (2.3 \AA) than Ir-O bond lengths.
However, quantitative analysis beyond this simple discussion is outside the scope of this paper.

The energy scale of the t$_{2g}$-e$_g$ $d-d$ excitation is closely related to the difference in the resonance energy of the $d-d$ excitations (Fig.~\ref{fig:res}) as discussed in the previous subsection. Both measure roughly the same energy scale $\sim 10Dq$, but there is a quantitative difference. The RIXS intermediate state (XAS final state) contains a core hole, which pulls the valence energy level down. All energy levels experience similar shifts, but because of this effect, the energy scale could be renormalized to a slightly smaller value. On the other hand, the RIXS final state does not have a core hole and therefore does not suffer from the core-hole effect. Another difference is that the x-ray absorption spectroscopy measures the splitting between the {\em empty} t$_{2g}$ states and the {\em empty} e$_g$ states, while RIXS measures the excitation energy between the {\em filled} t$_{2g}$ states and the {\em empty} e$_g$ states. Since the t$_{2g}$ states are only partially filled, the energy splitting measured with RIXS is larger than that measured with XAS.


Since the intra-t$_{2g}$ excitations below 1~eV in various Ir$^{4+}$ compounds have been investigated in detail in earlier studies, we will not go into detailed discussions about these excitations. However, we would like to point out that the RIXS spectrum of $\rm Sr_2YIrO_6$, in which Ir$^{5+}$ takes on $5d^4$ electronic configuration, is very different from those of Ir$^{4+}$ materials. The peak positions can be explained using the atomic model including both spin-orbit coupling and Hund's coupling as shown in Ref.~\onlinecite{Yuan2017} However, we observe extremely sharp peak widths in this case. In fact, these peaks are resolution limited, as one can easily compare their widths with the elastic peak width. As discussed in Sec.IIB, the narrow peak widths reflect the fact that Ir in this double perovskite compound is isolated from other Ir ions due to the intervening non-magnetic Y ions. The narrow peak widths, therefore, reflects negligible hybridization and very narrow Ir d bandwidths. On the other hand, the broad width of the intra-t$_{2g}$ excitations in IrO$_2$ reflects the wide metallic bandwidth in this compounds.\cite{Kahk2014}

\section{Conclusions}

We presented RXES and RIXS spectra of iridate samples with various structures, oxidation states, and transport properties. We discussed how quantitative and detailed information about unoccupied states can be obtained using the RXES map. One way is to use high energy resolution of the analyzer to obtain HERFD-XAS spectrum, which overcomes the core-hole lifetime broadening. We also illustrated one can obtain quasi $M_4$-edge absorption spectrum using hard x-rays, which could be useful for studying materials under various sample environments (e.g., diamond anvil cells, magnets, etc.).

Iridium $L_3$ edge RIXS spectrum is essentially dominated by electronic transitions between various 5d states, but actual spectra are extremely diverse.  Our comprehensive examination reveals a few general points to take away from RIXS studies.

{\em Resonance condition:} Because of the short core-hole lifetime, the RIXS resonance condition is simplified for $d-d$ excitations in iridates. The strong t$_{2g}-$e$_g$ transition intensity is resonantly enhanced near the peak of x-ray absorption, while the spin-orbit excitations and magnons at low energy ($< 1$~eV) resonate at an incident energy about 3~eV below the absorption peak. This information is quite helpful for choosing incident energy for future experiments. Our data also demonstrates that the ultra-short core-hole lifetime approximation works very well in the insulating iridates, facilitating comparison with theoretical calculations.\cite{UCL-EPL}

{\em Bandwidth:} The peak width is the quantity that varies the most among the iridate compounds studied here. Iridium ions in $\rm Sr_2YIrO_6$ are separated by another non-magnetic ion, which makes the transitions between the spin-orbit split states well-defined atomic-like transitions. The RIXS spectra show sharp resolution limited peaks in this case. On the other hand, metallic $\rm IrO_2$, the RIXS peaks should be considered as interband transitions, and the peak width represents the bandwidth. In all iridates the transitions between the occupied t$_{2g}$ states to unoccupied e$_g$ states tend to be broad, presumably reflecting stronger hybridization between e$_g$ orbitals and oxygen $p$ orbitals.

\begin{acknowledgements}
Research at the University of Toronto was supported by the Natural
Sciences and Engineering Research Council of Canada through
Discovery Grant and Research Tools and Instruments Grant. G.C. acknowledges NSF support via grant DMR-1712101.
This research used resources of the Advanced Photon Source, a U.S. Department of Energy
(DOE) Office of Science User Facility operated for the DOE Office of Science by Argonne
National Laboratory under Contract No. DE-AC02-06CH11357.

\end{acknowledgements}


\begin{thebibliography}{79}%
\makeatletter
\providecommand \@ifxundefined [1]{%
 \@ifx{#1\undefined}
}%
\providecommand \@ifnum [1]{%
 \ifnum #1\expandafter \@firstoftwo
 \else \expandafter \@secondoftwo
 \fi
}%
\providecommand \@ifx [1]{%
 \ifx #1\expandafter \@firstoftwo
 \else \expandafter \@secondoftwo
 \fi
}%
\providecommand \natexlab [1]{#1}%
\providecommand \enquote  [1]{``#1''}%
\providecommand \bibnamefont  [1]{#1}%
\providecommand \bibfnamefont [1]{#1}%
\providecommand \citenamefont [1]{#1}%
\providecommand \href@noop [0]{\@secondoftwo}%
\providecommand \href [0]{\begingroup \@sanitize@url \@href}%
\providecommand \@href[1]{\@@startlink{#1}\@@href}%
\providecommand \@@href[1]{\endgroup#1\@@endlink}%
\providecommand \@sanitize@url [0]{\catcode `\\12\catcode `\$12\catcode
  `\&12\catcode `\#12\catcode `\^12\catcode `\_12\catcode `\%12\relax}%
\providecommand \@@startlink[1]{}%
\providecommand \@@endlink[0]{}%
\providecommand \url  [0]{\begingroup\@sanitize@url \@url }%
\providecommand \@url [1]{\endgroup\@href {#1}{\urlprefix }}%
\providecommand \urlprefix  [0]{URL }%
\providecommand \Eprint [0]{\href }%
\providecommand \doibase [0]{http://dx.doi.org/}%
\providecommand \selectlanguage [0]{\@gobble}%
\providecommand \bibinfo  [0]{\@secondoftwo}%
\providecommand \bibfield  [0]{\@secondoftwo}%
\providecommand \translation [1]{[#1]}%
\providecommand \BibitemOpen [0]{}%
\providecommand \bibitemStop [0]{}%
\providecommand \bibitemNoStop [0]{.\EOS\space}%
\providecommand \EOS [0]{\spacefactor3000\relax}%
\providecommand \BibitemShut  [1]{\csname bibitem#1\endcsname}%
\let\auto@bib@innerbib\@empty
\bibitem [{\citenamefont {Kotani}\ and\ \citenamefont
  {Shin}(2001)}]{Kotani2001}%
  \BibitemOpen
  \bibfield  {author} {\bibinfo {author} {\bibfnamefont {A.}~\bibnamefont
  {Kotani}}\ and\ \bibinfo {author} {\bibfnamefont {S.}~\bibnamefont {Shin}},\
  }\href {\doibase 10.1103/RevModPhys.73.203} {\bibfield  {journal} {\bibinfo
  {journal} {Rev. Mod. Phys.}\ }\textbf {\bibinfo {volume} {73}},\ \bibinfo
  {pages} {203} (\bibinfo {year} {2001})}\BibitemShut {NoStop}%
\bibitem [{\citenamefont {Rueff}\ and\ \citenamefont
  {Shukla}(2010)}]{Rueff2010}%
  \BibitemOpen
  \bibfield  {author} {\bibinfo {author} {\bibfnamefont {J.-P.}\ \bibnamefont
  {Rueff}}\ and\ \bibinfo {author} {\bibfnamefont {A.}~\bibnamefont {Shukla}},\
  }\href {\doibase 10.1103/RevModPhys.82.847} {\bibfield  {journal} {\bibinfo
  {journal} {Rev. Mod. Phys.}\ }\textbf {\bibinfo {volume} {82}},\ \bibinfo
  {pages} {847} (\bibinfo {year} {2010})}\BibitemShut {NoStop}%
\bibitem [{\citenamefont {Ament}\ \emph {et~al.}(2011)\citenamefont {Ament},
  \citenamefont {van Veenendaal}, \citenamefont {Devereaux}, \citenamefont
  {Hill},\ and\ \citenamefont {van~den Brink}}]{Ament2011}%
  \BibitemOpen
  \bibfield  {author} {\bibinfo {author} {\bibfnamefont {L.~J.~P.}\
  \bibnamefont {Ament}}, \bibinfo {author} {\bibfnamefont {M.}~\bibnamefont
  {van Veenendaal}}, \bibinfo {author} {\bibfnamefont {T.~P.}\ \bibnamefont
  {Devereaux}}, \bibinfo {author} {\bibfnamefont {J.~P.}\ \bibnamefont {Hill}},
  \ and\ \bibinfo {author} {\bibfnamefont {J.}~\bibnamefont {van~den Brink}},\
  }\href {\doibase 10.1103/RevModPhys.83.705} {\bibfield  {journal} {\bibinfo
  {journal} {Rev. Mod. Phys.}\ }\textbf {\bibinfo {volume} {83}},\ \bibinfo
  {pages} {705} (\bibinfo {year} {2011})}\BibitemShut {NoStop}%
\bibitem [{\citenamefont {Sch{\"u}lke}(2007)}]{Schulke_book}%
  \BibitemOpen
  \bibfield  {author} {\bibinfo {author} {\bibfnamefont {W.}~\bibnamefont
  {Sch{\"u}lke}},\ }\href@noop {} {\emph {\bibinfo {title} {Electron dynamics
  by inelastic X-ray scattering}}}\ (\bibinfo  {publisher} {Oxford University
  Press, Oxford},\ \bibinfo {year} {2007})\BibitemShut {NoStop}%
\bibitem [{\citenamefont {De~Groot}\ and\ \citenamefont
  {Kotani}(2008)}]{deGroot_book}%
  \BibitemOpen
  \bibfield  {author} {\bibinfo {author} {\bibfnamefont {F.}~\bibnamefont
  {De~Groot}}\ and\ \bibinfo {author} {\bibfnamefont {A.}~\bibnamefont
  {Kotani}},\ }\href@noop {} {\emph {\bibinfo {title} {Core level spectroscopy
  of solids}}}\ (\bibinfo  {publisher} {CRC press},\ \bibinfo {year}
  {2008})\BibitemShut {NoStop}%
\bibitem [{\citenamefont {de~Groot}\ \emph {et~al.}(1998)\citenamefont
  {de~Groot}, \citenamefont {Kuiper},\ and\ \citenamefont
  {Sawatzky}}]{deGroot1998}%
  \BibitemOpen
  \bibfield  {author} {\bibinfo {author} {\bibfnamefont {F.~M.~F.}\
  \bibnamefont {de~Groot}}, \bibinfo {author} {\bibfnamefont {P.}~\bibnamefont
  {Kuiper}}, \ and\ \bibinfo {author} {\bibfnamefont {G.~A.}\ \bibnamefont
  {Sawatzky}},\ }\href {\doibase 10.1103/PhysRevB.57.14584} {\bibfield
  {journal} {\bibinfo  {journal} {Phys. Rev. B}\ }\textbf {\bibinfo {volume}
  {57}},\ \bibinfo {pages} {14584} (\bibinfo {year} {1998})}\BibitemShut
  {NoStop}%
\bibitem [{\citenamefont {Kuiper}\ \emph {et~al.}(1998)\citenamefont {Kuiper},
  \citenamefont {Guo}, \citenamefont {S\aa{}the}, \citenamefont {Duda},
  \citenamefont {Nordgren}, \citenamefont {Pothuizen}, \citenamefont
  {de~Groot},\ and\ \citenamefont {Sawatzky}}]{Kuiper1998}%
  \BibitemOpen
  \bibfield  {author} {\bibinfo {author} {\bibfnamefont {P.}~\bibnamefont
  {Kuiper}}, \bibinfo {author} {\bibfnamefont {J.-H.}\ \bibnamefont {Guo}},
  \bibinfo {author} {\bibfnamefont {C.}~\bibnamefont {S\aa{}the}}, \bibinfo
  {author} {\bibfnamefont {L.-C.}\ \bibnamefont {Duda}}, \bibinfo {author}
  {\bibfnamefont {J.}~\bibnamefont {Nordgren}}, \bibinfo {author}
  {\bibfnamefont {J.~J.~M.}\ \bibnamefont {Pothuizen}}, \bibinfo {author}
  {\bibfnamefont {F.~M.~F.}\ \bibnamefont {de~Groot}}, \ and\ \bibinfo {author}
  {\bibfnamefont {G.~A.}\ \bibnamefont {Sawatzky}},\ }\href {\doibase
  10.1103/PhysRevLett.80.5204} {\bibfield  {journal} {\bibinfo  {journal}
  {Phys. Rev. Lett.}\ }\textbf {\bibinfo {volume} {80}},\ \bibinfo {pages}
  {5204} (\bibinfo {year} {1998})}\BibitemShut {NoStop}%
\bibitem [{\citenamefont {Chiuzb\ifmmode~\u{a}\else \u{a}\fi{}ian}\ \emph
  {et~al.}(2005)\citenamefont {Chiuzb\ifmmode~\u{a}\else \u{a}\fi{}ian},
  \citenamefont {Ghiringhelli}, \citenamefont {Dallera}, \citenamefont
  {Grioni}, \citenamefont {Amann}, \citenamefont {Wang}, \citenamefont
  {Braicovich},\ and\ \citenamefont {Patthey}}]{Chiuzbaian2005}%
  \BibitemOpen
  \bibfield  {author} {\bibinfo {author} {\bibfnamefont {S.~G.}\ \bibnamefont
  {Chiuzb\ifmmode~\u{a}\else \u{a}\fi{}ian}}, \bibinfo {author} {\bibfnamefont
  {G.}~\bibnamefont {Ghiringhelli}}, \bibinfo {author} {\bibfnamefont
  {C.}~\bibnamefont {Dallera}}, \bibinfo {author} {\bibfnamefont
  {M.}~\bibnamefont {Grioni}}, \bibinfo {author} {\bibfnamefont
  {P.}~\bibnamefont {Amann}}, \bibinfo {author} {\bibfnamefont
  {X.}~\bibnamefont {Wang}}, \bibinfo {author} {\bibfnamefont {L.}~\bibnamefont
  {Braicovich}}, \ and\ \bibinfo {author} {\bibfnamefont {L.}~\bibnamefont
  {Patthey}},\ }\href {\doibase 10.1103/PhysRevLett.95.197402} {\bibfield
  {journal} {\bibinfo  {journal} {Phys. Rev. Lett.}\ }\textbf {\bibinfo
  {volume} {95}},\ \bibinfo {pages} {197402} (\bibinfo {year}
  {2005})}\BibitemShut {NoStop}%
\bibitem [{\citenamefont {Hill}\ \emph {et~al.}(2008)\citenamefont {Hill},
  \citenamefont {Blumberg}, \citenamefont {Kim}, \citenamefont {Ellis},
  \citenamefont {Wakimoto}, \citenamefont {Birgeneau}, \citenamefont {Komiya},
  \citenamefont {Ando}, \citenamefont {Liang}, \citenamefont {Greene},
  \citenamefont {Casa},\ and\ \citenamefont {Gog}}]{Hill2008}%
  \BibitemOpen
  \bibfield  {author} {\bibinfo {author} {\bibfnamefont {J.~P.}\ \bibnamefont
  {Hill}}, \bibinfo {author} {\bibfnamefont {G.}~\bibnamefont {Blumberg}},
  \bibinfo {author} {\bibfnamefont {Y.-J.}\ \bibnamefont {Kim}}, \bibinfo
  {author} {\bibfnamefont {D.~S.}\ \bibnamefont {Ellis}}, \bibinfo {author}
  {\bibfnamefont {S.}~\bibnamefont {Wakimoto}}, \bibinfo {author}
  {\bibfnamefont {R.~J.}\ \bibnamefont {Birgeneau}}, \bibinfo {author}
  {\bibfnamefont {S.}~\bibnamefont {Komiya}}, \bibinfo {author} {\bibfnamefont
  {Y.}~\bibnamefont {Ando}}, \bibinfo {author} {\bibfnamefont {B.}~\bibnamefont
  {Liang}}, \bibinfo {author} {\bibfnamefont {R.~L.}\ \bibnamefont {Greene}},
  \bibinfo {author} {\bibfnamefont {D.}~\bibnamefont {Casa}}, \ and\ \bibinfo
  {author} {\bibfnamefont {T.}~\bibnamefont {Gog}},\ }\href {\doibase
  10.1103/PhysRevLett.100.097001} {\bibfield  {journal} {\bibinfo  {journal}
  {Phys. Rev. Lett.}\ }\textbf {\bibinfo {volume} {100}},\ \bibinfo {pages}
  {097001} (\bibinfo {year} {2008})}\BibitemShut {NoStop}%
\bibitem [{\citenamefont {Ellis}\ \emph {et~al.}(2010)\citenamefont {Ellis},
  \citenamefont {Kim}, \citenamefont {Hill}, \citenamefont {Wakimoto},
  \citenamefont {Birgeneau}, \citenamefont {Shvyd'ko}, \citenamefont {Casa},
  \citenamefont {Gog}, \citenamefont {Ishii}, \citenamefont {Ikeuchi},
  \citenamefont {Paramekanti},\ and\ \citenamefont {Kim}}]{Ellis2010}%
  \BibitemOpen
  \bibfield  {author} {\bibinfo {author} {\bibfnamefont {D.~S.}\ \bibnamefont
  {Ellis}}, \bibinfo {author} {\bibfnamefont {J.}~\bibnamefont {Kim}}, \bibinfo
  {author} {\bibfnamefont {J.~P.}\ \bibnamefont {Hill}}, \bibinfo {author}
  {\bibfnamefont {S.}~\bibnamefont {Wakimoto}}, \bibinfo {author}
  {\bibfnamefont {R.~J.}\ \bibnamefont {Birgeneau}}, \bibinfo {author}
  {\bibfnamefont {Y.}~\bibnamefont {Shvyd'ko}}, \bibinfo {author}
  {\bibfnamefont {D.}~\bibnamefont {Casa}}, \bibinfo {author} {\bibfnamefont
  {T.}~\bibnamefont {Gog}}, \bibinfo {author} {\bibfnamefont {K.}~\bibnamefont
  {Ishii}}, \bibinfo {author} {\bibfnamefont {K.}~\bibnamefont {Ikeuchi}},
  \bibinfo {author} {\bibfnamefont {A.}~\bibnamefont {Paramekanti}}, \ and\
  \bibinfo {author} {\bibfnamefont {Y.-J.}\ \bibnamefont {Kim}},\ }\href
  {\doibase 10.1103/PhysRevB.81.085124} {\bibfield  {journal} {\bibinfo
  {journal} {Phys. Rev. B}\ }\textbf {\bibinfo {volume} {81}},\ \bibinfo
  {pages} {085124} (\bibinfo {year} {2010})}\BibitemShut {NoStop}%
\bibitem [{\citenamefont {Braicovich}\ \emph {et~al.}(2009)\citenamefont
  {Braicovich}, \citenamefont {Ament}, \citenamefont {Bisogni}, \citenamefont
  {Forte}, \citenamefont {Aruta}, \citenamefont {Balestrino}, \citenamefont
  {Brookes}, \citenamefont {De~Luca}, \citenamefont {Medaglia}, \citenamefont
  {Granozio}, \citenamefont {Radovic}, \citenamefont {Salluzzo}, \citenamefont
  {van~den Brink},\ and\ \citenamefont {Ghiringhelli}}]{Braicovich2009}%
  \BibitemOpen
  \bibfield  {author} {\bibinfo {author} {\bibfnamefont {L.}~\bibnamefont
  {Braicovich}}, \bibinfo {author} {\bibfnamefont {L.~J.~P.}\ \bibnamefont
  {Ament}}, \bibinfo {author} {\bibfnamefont {V.}~\bibnamefont {Bisogni}},
  \bibinfo {author} {\bibfnamefont {F.}~\bibnamefont {Forte}}, \bibinfo
  {author} {\bibfnamefont {C.}~\bibnamefont {Aruta}}, \bibinfo {author}
  {\bibfnamefont {G.}~\bibnamefont {Balestrino}}, \bibinfo {author}
  {\bibfnamefont {N.~B.}\ \bibnamefont {Brookes}}, \bibinfo {author}
  {\bibfnamefont {G.~M.}\ \bibnamefont {De~Luca}}, \bibinfo {author}
  {\bibfnamefont {P.~G.}\ \bibnamefont {Medaglia}}, \bibinfo {author}
  {\bibfnamefont {F.~M.}\ \bibnamefont {Granozio}}, \bibinfo {author}
  {\bibfnamefont {M.}~\bibnamefont {Radovic}}, \bibinfo {author} {\bibfnamefont
  {M.}~\bibnamefont {Salluzzo}}, \bibinfo {author} {\bibfnamefont
  {J.}~\bibnamefont {van~den Brink}}, \ and\ \bibinfo {author} {\bibfnamefont
  {G.}~\bibnamefont {Ghiringhelli}},\ }\href {\doibase
  10.1103/PhysRevLett.102.167401} {\bibfield  {journal} {\bibinfo  {journal}
  {Phys. Rev. Lett.}\ }\textbf {\bibinfo {volume} {102}},\ \bibinfo {pages}
  {167401} (\bibinfo {year} {2009})}\BibitemShut {NoStop}%
\bibitem [{\citenamefont {Ament}\ \emph {et~al.}(2009)\citenamefont {Ament},
  \citenamefont {Ghiringhelli}, \citenamefont {Sala}, \citenamefont
  {Braicovich},\ and\ \citenamefont {van~den Brink}}]{Ament2009}%
  \BibitemOpen
  \bibfield  {author} {\bibinfo {author} {\bibfnamefont {L.~J.~P.}\
  \bibnamefont {Ament}}, \bibinfo {author} {\bibfnamefont {G.}~\bibnamefont
  {Ghiringhelli}}, \bibinfo {author} {\bibfnamefont {M.~M.}\ \bibnamefont
  {Sala}}, \bibinfo {author} {\bibfnamefont {L.}~\bibnamefont {Braicovich}}, \
  and\ \bibinfo {author} {\bibfnamefont {J.}~\bibnamefont {van~den Brink}},\
  }\href {\doibase 10.1103/PhysRevLett.103.117003} {\bibfield  {journal}
  {\bibinfo  {journal} {Phys. Rev. Lett.}\ }\textbf {\bibinfo {volume} {103}},\
  \bibinfo {pages} {117003} (\bibinfo {year} {2009})}\BibitemShut {NoStop}%
\bibitem [{\citenamefont {Haverkort}(2010)}]{Haverkort2010}%
  \BibitemOpen
  \bibfield  {author} {\bibinfo {author} {\bibfnamefont {M.~W.}\ \bibnamefont
  {Haverkort}},\ }\href {\doibase 10.1103/PhysRevLett.105.167404} {\bibfield
  {journal} {\bibinfo  {journal} {Phys. Rev. Lett.}\ }\textbf {\bibinfo
  {volume} {105}},\ \bibinfo {pages} {167404} (\bibinfo {year}
  {2010})}\BibitemShut {NoStop}%
\bibitem [{\citenamefont {Schlappa}\ \emph {et~al.}(2009)\citenamefont
  {Schlappa}, \citenamefont {Schmitt}, \citenamefont {Vernay}, \citenamefont
  {Strocov}, \citenamefont {Ilakovac}, \citenamefont {Thielemann},
  \citenamefont {R\o{}nnow}, \citenamefont {Vanishri}, \citenamefont
  {Piazzalunga}, \citenamefont {Wang}, \citenamefont {Braicovich},
  \citenamefont {Ghiringhelli}, \citenamefont {Marin}, \citenamefont {Mesot},
  \citenamefont {Delley},\ and\ \citenamefont {Patthey}}]{Schlappa2009}%
  \BibitemOpen
  \bibfield  {author} {\bibinfo {author} {\bibfnamefont {J.}~\bibnamefont
  {Schlappa}}, \bibinfo {author} {\bibfnamefont {T.}~\bibnamefont {Schmitt}},
  \bibinfo {author} {\bibfnamefont {F.}~\bibnamefont {Vernay}}, \bibinfo
  {author} {\bibfnamefont {V.~N.}\ \bibnamefont {Strocov}}, \bibinfo {author}
  {\bibfnamefont {V.}~\bibnamefont {Ilakovac}}, \bibinfo {author}
  {\bibfnamefont {B.}~\bibnamefont {Thielemann}}, \bibinfo {author}
  {\bibfnamefont {H.~M.}\ \bibnamefont {R\o{}nnow}}, \bibinfo {author}
  {\bibfnamefont {S.}~\bibnamefont {Vanishri}}, \bibinfo {author}
  {\bibfnamefont {A.}~\bibnamefont {Piazzalunga}}, \bibinfo {author}
  {\bibfnamefont {X.}~\bibnamefont {Wang}}, \bibinfo {author} {\bibfnamefont
  {L.}~\bibnamefont {Braicovich}}, \bibinfo {author} {\bibfnamefont
  {G.}~\bibnamefont {Ghiringhelli}}, \bibinfo {author} {\bibfnamefont
  {C.}~\bibnamefont {Marin}}, \bibinfo {author} {\bibfnamefont
  {J.}~\bibnamefont {Mesot}}, \bibinfo {author} {\bibfnamefont
  {B.}~\bibnamefont {Delley}}, \ and\ \bibinfo {author} {\bibfnamefont
  {L.}~\bibnamefont {Patthey}},\ }\href {\doibase
  10.1103/PhysRevLett.103.047401} {\bibfield  {journal} {\bibinfo  {journal}
  {Phys. Rev. Lett.}\ }\textbf {\bibinfo {volume} {103}},\ \bibinfo {pages}
  {047401} (\bibinfo {year} {2009})}\BibitemShut {NoStop}%
\bibitem [{\citenamefont {Le~Tacon}\ \emph {et~al.}(2011)\citenamefont
  {Le~Tacon}, \citenamefont {Ghiringhelli}, \citenamefont {Chaloupka},
  \citenamefont {Sala}, \citenamefont {Hinkov}, \citenamefont {Haverkort},
  \citenamefont {Minola}, \citenamefont {Bakr}, \citenamefont {Zhou},
  \citenamefont {Blanco-Canosa} \emph {et~al.}}]{LeTacon2011}%
  \BibitemOpen
  \bibfield  {author} {\bibinfo {author} {\bibfnamefont {M.}~\bibnamefont
  {Le~Tacon}}, \bibinfo {author} {\bibfnamefont {G.}~\bibnamefont
  {Ghiringhelli}}, \bibinfo {author} {\bibfnamefont {J.}~\bibnamefont
  {Chaloupka}}, \bibinfo {author} {\bibfnamefont {M.~M.}\ \bibnamefont {Sala}},
  \bibinfo {author} {\bibfnamefont {V.}~\bibnamefont {Hinkov}}, \bibinfo
  {author} {\bibfnamefont {M.}~\bibnamefont {Haverkort}}, \bibinfo {author}
  {\bibfnamefont {M.}~\bibnamefont {Minola}}, \bibinfo {author} {\bibfnamefont
  {M.}~\bibnamefont {Bakr}}, \bibinfo {author} {\bibfnamefont {K.}~\bibnamefont
  {Zhou}}, \bibinfo {author} {\bibfnamefont {S.}~\bibnamefont {Blanco-Canosa}},
   \emph {et~al.},\ }\href@noop {} {\bibfield  {journal} {\bibinfo  {journal}
  {Nature Physics}\ }\textbf {\bibinfo {volume} {7}},\ \bibinfo {pages} {725}
  (\bibinfo {year} {2011})}\BibitemShut {NoStop}%
\bibitem [{\citenamefont {Schlappa}\ \emph {et~al.}(2012)\citenamefont
  {Schlappa}, \citenamefont {Wohlfeld}, \citenamefont {Zhou}, \citenamefont
  {Mourigal}, \citenamefont {Haverkort}, \citenamefont {Strocov}, \citenamefont
  {Hozoi}, \citenamefont {Monney}, \citenamefont {Nishimoto}, \citenamefont
  {Singh} \emph {et~al.}}]{Schlappa2012}%
  \BibitemOpen
  \bibfield  {author} {\bibinfo {author} {\bibfnamefont {J.}~\bibnamefont
  {Schlappa}}, \bibinfo {author} {\bibfnamefont {K.}~\bibnamefont {Wohlfeld}},
  \bibinfo {author} {\bibfnamefont {K.}~\bibnamefont {Zhou}}, \bibinfo {author}
  {\bibfnamefont {M.}~\bibnamefont {Mourigal}}, \bibinfo {author}
  {\bibfnamefont {M.}~\bibnamefont {Haverkort}}, \bibinfo {author}
  {\bibfnamefont {V.}~\bibnamefont {Strocov}}, \bibinfo {author} {\bibfnamefont
  {L.}~\bibnamefont {Hozoi}}, \bibinfo {author} {\bibfnamefont
  {C.}~\bibnamefont {Monney}}, \bibinfo {author} {\bibfnamefont
  {S.}~\bibnamefont {Nishimoto}}, \bibinfo {author} {\bibfnamefont
  {S.}~\bibnamefont {Singh}},  \emph {et~al.},\ }\href@noop {} {\bibfield
  {journal} {\bibinfo  {journal} {Nature}\ }\textbf {\bibinfo {volume} {485}},\
  \bibinfo {pages} {82} (\bibinfo {year} {2012})}\BibitemShut {NoStop}%
\bibitem [{\citenamefont {Dean}\ \emph {et~al.}(2012)\citenamefont {Dean},
  \citenamefont {Springell}, \citenamefont {Monney}, \citenamefont {Zhou},
  \citenamefont {Pereiro}, \citenamefont {Bo{\v{z}}ovi{\'c}}, \citenamefont
  {Dalla~Piazza}, \citenamefont {R{\o}nnow}, \citenamefont {Morenzoni},
  \citenamefont {Van Den~Brink} \emph {et~al.}}]{Dean2012}%
  \BibitemOpen
  \bibfield  {author} {\bibinfo {author} {\bibfnamefont {M.~P.~M.}\
  \bibnamefont {Dean}}, \bibinfo {author} {\bibfnamefont {R.}~\bibnamefont
  {Springell}}, \bibinfo {author} {\bibfnamefont {C.}~\bibnamefont {Monney}},
  \bibinfo {author} {\bibfnamefont {K.}~\bibnamefont {Zhou}}, \bibinfo {author}
  {\bibfnamefont {J.}~\bibnamefont {Pereiro}}, \bibinfo {author} {\bibfnamefont
  {I.}~\bibnamefont {Bo{\v{z}}ovi{\'c}}}, \bibinfo {author} {\bibfnamefont
  {B.}~\bibnamefont {Dalla~Piazza}}, \bibinfo {author} {\bibfnamefont
  {H.}~\bibnamefont {R{\o}nnow}}, \bibinfo {author} {\bibfnamefont
  {E.}~\bibnamefont {Morenzoni}}, \bibinfo {author} {\bibfnamefont
  {J.}~\bibnamefont {Van Den~Brink}},  \emph {et~al.},\ }\href@noop {}
  {\bibfield  {journal} {\bibinfo  {journal} {Nature materials}\ }\textbf
  {\bibinfo {volume} {11}},\ \bibinfo {pages} {850} (\bibinfo {year}
  {2012})}\BibitemShut {NoStop}%
\bibitem [{\citenamefont {Dean}\ \emph
  {et~al.}(2013{\natexlab{a}})\citenamefont {Dean}, \citenamefont {Dellea},
  \citenamefont {Springell}, \citenamefont {Yakhou-Harris}, \citenamefont
  {Kummer}, \citenamefont {Brookes}, \citenamefont {Liu}, \citenamefont {Sun},
  \citenamefont {Strle}, \citenamefont {Schmitt} \emph {et~al.}}]{Dean2013a}%
  \BibitemOpen
  \bibfield  {author} {\bibinfo {author} {\bibfnamefont {M.~P.~M.}\
  \bibnamefont {Dean}}, \bibinfo {author} {\bibfnamefont {G.}~\bibnamefont
  {Dellea}}, \bibinfo {author} {\bibfnamefont {R.}~\bibnamefont {Springell}},
  \bibinfo {author} {\bibfnamefont {F.}~\bibnamefont {Yakhou-Harris}}, \bibinfo
  {author} {\bibfnamefont {K.}~\bibnamefont {Kummer}}, \bibinfo {author}
  {\bibfnamefont {N.}~\bibnamefont {Brookes}}, \bibinfo {author} {\bibfnamefont
  {X.}~\bibnamefont {Liu}}, \bibinfo {author} {\bibfnamefont {Y.}~\bibnamefont
  {Sun}}, \bibinfo {author} {\bibfnamefont {J.}~\bibnamefont {Strle}}, \bibinfo
  {author} {\bibfnamefont {T.}~\bibnamefont {Schmitt}},  \emph {et~al.},\
  }\href@noop {} {\bibfield  {journal} {\bibinfo  {journal} {Nature materials}\
  }\textbf {\bibinfo {volume} {12}},\ \bibinfo {pages} {1019} (\bibinfo {year}
  {2013}{\natexlab{a}})}\BibitemShut {NoStop}%
\bibitem [{\citenamefont {Dean}\ \emph
  {et~al.}(2013{\natexlab{b}})\citenamefont {Dean}, \citenamefont {James},
  \citenamefont {Springell}, \citenamefont {Liu}, \citenamefont {Monney},
  \citenamefont {Zhou}, \citenamefont {Konik}, \citenamefont {Wen},
  \citenamefont {Xu}, \citenamefont {Gu}, \citenamefont {Strocov},
  \citenamefont {Schmitt},\ and\ \citenamefont {Hill}}]{Dean2013b}%
  \BibitemOpen
  \bibfield  {author} {\bibinfo {author} {\bibfnamefont {M.~P.~M.}\
  \bibnamefont {Dean}}, \bibinfo {author} {\bibfnamefont {A.~J.~A.}\
  \bibnamefont {James}}, \bibinfo {author} {\bibfnamefont {R.~S.}\ \bibnamefont
  {Springell}}, \bibinfo {author} {\bibfnamefont {X.}~\bibnamefont {Liu}},
  \bibinfo {author} {\bibfnamefont {C.}~\bibnamefont {Monney}}, \bibinfo
  {author} {\bibfnamefont {K.~J.}\ \bibnamefont {Zhou}}, \bibinfo {author}
  {\bibfnamefont {R.~M.}\ \bibnamefont {Konik}}, \bibinfo {author}
  {\bibfnamefont {J.~S.}\ \bibnamefont {Wen}}, \bibinfo {author} {\bibfnamefont
  {Z.~J.}\ \bibnamefont {Xu}}, \bibinfo {author} {\bibfnamefont {G.~D.}\
  \bibnamefont {Gu}}, \bibinfo {author} {\bibfnamefont {V.~N.}\ \bibnamefont
  {Strocov}}, \bibinfo {author} {\bibfnamefont {T.}~\bibnamefont {Schmitt}}, \
  and\ \bibinfo {author} {\bibfnamefont {J.~P.}\ \bibnamefont {Hill}},\ }\href
  {\doibase 10.1103/PhysRevLett.110.147001} {\bibfield  {journal} {\bibinfo
  {journal} {Phys. Rev. Lett.}\ }\textbf {\bibinfo {volume} {110}},\ \bibinfo
  {pages} {147001} (\bibinfo {year} {2013}{\natexlab{b}})}\BibitemShut
  {NoStop}%
\bibitem [{\citenamefont {Dean}(2015)}]{Dean2015}%
  \BibitemOpen
  \bibfield  {author} {\bibinfo {author} {\bibfnamefont {M.}~\bibnamefont
  {Dean}},\ }\href {\doibase http://dx.doi.org/10.1016/j.jmmm.2014.03.057}
  {\bibfield  {journal} {\bibinfo  {journal} {Journal of Magnetism and Magnetic
  Materials}\ }\textbf {\bibinfo {volume} {376}},\ \bibinfo {pages} {3 }
  (\bibinfo {year} {2015})},\ \bibinfo {note} {pseudogap, Superconductivity and
  Magnetism}\BibitemShut {NoStop}%
\bibitem [{\citenamefont {Ishii}\ \emph {et~al.}(2011)\citenamefont {Ishii},
  \citenamefont {Jarrige}, \citenamefont {Yoshida}, \citenamefont {Ikeuchi},
  \citenamefont {Mizuki}, \citenamefont {Ohashi}, \citenamefont {Takayama},
  \citenamefont {Matsuno},\ and\ \citenamefont {Takagi}}]{Ishii2011}%
  \BibitemOpen
  \bibfield  {author} {\bibinfo {author} {\bibfnamefont {K.}~\bibnamefont
  {Ishii}}, \bibinfo {author} {\bibfnamefont {I.}~\bibnamefont {Jarrige}},
  \bibinfo {author} {\bibfnamefont {M.}~\bibnamefont {Yoshida}}, \bibinfo
  {author} {\bibfnamefont {K.}~\bibnamefont {Ikeuchi}}, \bibinfo {author}
  {\bibfnamefont {J.}~\bibnamefont {Mizuki}}, \bibinfo {author} {\bibfnamefont
  {K.}~\bibnamefont {Ohashi}}, \bibinfo {author} {\bibfnamefont
  {T.}~\bibnamefont {Takayama}}, \bibinfo {author} {\bibfnamefont
  {J.}~\bibnamefont {Matsuno}}, \ and\ \bibinfo {author} {\bibfnamefont
  {H.}~\bibnamefont {Takagi}},\ }\href {\doibase 10.1103/PhysRevB.83.115121}
  {\bibfield  {journal} {\bibinfo  {journal} {Phys. Rev. B}\ }\textbf {\bibinfo
  {volume} {83}},\ \bibinfo {pages} {115121} (\bibinfo {year}
  {2011})}\BibitemShut {NoStop}%
\bibitem [{\citenamefont {Kim}\ \emph {et~al.}(2012{\natexlab{a}})\citenamefont
  {Kim}, \citenamefont {Casa}, \citenamefont {Upton}, \citenamefont {Gog},
  \citenamefont {Kim}, \citenamefont {Mitchell}, \citenamefont {van
  Veenendaal}, \citenamefont {Daghofer}, \citenamefont {van~den Brink},
  \citenamefont {Khaliullin},\ and\ \citenamefont {Kim}}]{JKim2012a}%
  \BibitemOpen
  \bibfield  {author} {\bibinfo {author} {\bibfnamefont {J.}~\bibnamefont
  {Kim}}, \bibinfo {author} {\bibfnamefont {D.}~\bibnamefont {Casa}}, \bibinfo
  {author} {\bibfnamefont {M.~H.}\ \bibnamefont {Upton}}, \bibinfo {author}
  {\bibfnamefont {T.}~\bibnamefont {Gog}}, \bibinfo {author} {\bibfnamefont
  {Y.-J.}\ \bibnamefont {Kim}}, \bibinfo {author} {\bibfnamefont {J.~F.}\
  \bibnamefont {Mitchell}}, \bibinfo {author} {\bibfnamefont {M.}~\bibnamefont
  {van Veenendaal}}, \bibinfo {author} {\bibfnamefont {M.}~\bibnamefont
  {Daghofer}}, \bibinfo {author} {\bibfnamefont {J.}~\bibnamefont {van~den
  Brink}}, \bibinfo {author} {\bibfnamefont {G.}~\bibnamefont {Khaliullin}}, \
  and\ \bibinfo {author} {\bibfnamefont {B.~J.}\ \bibnamefont {Kim}},\ }\href
  {\doibase 10.1103/PhysRevLett.108.177003} {\bibfield  {journal} {\bibinfo
  {journal} {Phys. Rev. Lett.}\ }\textbf {\bibinfo {volume} {108}},\ \bibinfo
  {pages} {177003} (\bibinfo {year} {2012}{\natexlab{a}})}\BibitemShut
  {NoStop}%
\bibitem [{\citenamefont {Dhital}\ \emph {et~al.}(2013)\citenamefont {Dhital},
  \citenamefont {Hogan}, \citenamefont {Yamani}, \citenamefont {de~la Cruz},
  \citenamefont {Chen}, \citenamefont {Khadka}, \citenamefont {Ren},\ and\
  \citenamefont {Wilson}}]{Dhital2013}%
  \BibitemOpen
  \bibfield  {author} {\bibinfo {author} {\bibfnamefont {C.}~\bibnamefont
  {Dhital}}, \bibinfo {author} {\bibfnamefont {T.}~\bibnamefont {Hogan}},
  \bibinfo {author} {\bibfnamefont {Z.}~\bibnamefont {Yamani}}, \bibinfo
  {author} {\bibfnamefont {C.}~\bibnamefont {de~la Cruz}}, \bibinfo {author}
  {\bibfnamefont {X.}~\bibnamefont {Chen}}, \bibinfo {author} {\bibfnamefont
  {S.}~\bibnamefont {Khadka}}, \bibinfo {author} {\bibfnamefont
  {Z.}~\bibnamefont {Ren}}, \ and\ \bibinfo {author} {\bibfnamefont {S.~D.}\
  \bibnamefont {Wilson}},\ }\href {\doibase 10.1103/PhysRevB.87.144405}
  {\bibfield  {journal} {\bibinfo  {journal} {Phys. Rev. B}\ }\textbf {\bibinfo
  {volume} {87}},\ \bibinfo {pages} {144405} (\bibinfo {year}
  {2013})}\BibitemShut {NoStop}%
\bibitem [{\citenamefont {Ye}\ \emph {et~al.}(2012)\citenamefont {Ye},
  \citenamefont {Chi}, \citenamefont {Cao}, \citenamefont {Chakoumakos},
  \citenamefont {Fernandez-Baca}, \citenamefont {Custelcean}, \citenamefont
  {Qi}, \citenamefont {Korneta},\ and\ \citenamefont {Cao}}]{Ye2012}%
  \BibitemOpen
  \bibfield  {author} {\bibinfo {author} {\bibfnamefont {F.}~\bibnamefont
  {Ye}}, \bibinfo {author} {\bibfnamefont {S.}~\bibnamefont {Chi}}, \bibinfo
  {author} {\bibfnamefont {H.}~\bibnamefont {Cao}}, \bibinfo {author}
  {\bibfnamefont {B.~C.}\ \bibnamefont {Chakoumakos}}, \bibinfo {author}
  {\bibfnamefont {J.~A.}\ \bibnamefont {Fernandez-Baca}}, \bibinfo {author}
  {\bibfnamefont {R.}~\bibnamefont {Custelcean}}, \bibinfo {author}
  {\bibfnamefont {T.~F.}\ \bibnamefont {Qi}}, \bibinfo {author} {\bibfnamefont
  {O.~B.}\ \bibnamefont {Korneta}}, \ and\ \bibinfo {author} {\bibfnamefont
  {G.}~\bibnamefont {Cao}},\ }\href {\doibase 10.1103/PhysRevB.85.180403}
  {\bibfield  {journal} {\bibinfo  {journal} {Phys. Rev. B}\ }\textbf {\bibinfo
  {volume} {85}},\ \bibinfo {pages} {180403} (\bibinfo {year}
  {2012})}\BibitemShut {NoStop}%
\bibitem [{\citenamefont {Choi}\ \emph {et~al.}(2012)\citenamefont {Choi},
  \citenamefont {Coldea}, \citenamefont {Kolmogorov}, \citenamefont
  {Lancaster}, \citenamefont {Mazin}, \citenamefont {Blundell}, \citenamefont
  {Radaelli}, \citenamefont {Singh}, \citenamefont {Gegenwart}, \citenamefont
  {Choi}, \citenamefont {Cheong}, \citenamefont {Baker}, \citenamefont
  {Stock},\ and\ \citenamefont {Taylor}}]{Choi2012}%
  \BibitemOpen
  \bibfield  {author} {\bibinfo {author} {\bibfnamefont {S.~K.}\ \bibnamefont
  {Choi}}, \bibinfo {author} {\bibfnamefont {R.}~\bibnamefont {Coldea}},
  \bibinfo {author} {\bibfnamefont {A.~N.}\ \bibnamefont {Kolmogorov}},
  \bibinfo {author} {\bibfnamefont {T.}~\bibnamefont {Lancaster}}, \bibinfo
  {author} {\bibfnamefont {I.~I.}\ \bibnamefont {Mazin}}, \bibinfo {author}
  {\bibfnamefont {S.~J.}\ \bibnamefont {Blundell}}, \bibinfo {author}
  {\bibfnamefont {P.~G.}\ \bibnamefont {Radaelli}}, \bibinfo {author}
  {\bibfnamefont {Y.}~\bibnamefont {Singh}}, \bibinfo {author} {\bibfnamefont
  {P.}~\bibnamefont {Gegenwart}}, \bibinfo {author} {\bibfnamefont {K.~R.}\
  \bibnamefont {Choi}}, \bibinfo {author} {\bibfnamefont {S.-W.}\ \bibnamefont
  {Cheong}}, \bibinfo {author} {\bibfnamefont {P.~J.}\ \bibnamefont {Baker}},
  \bibinfo {author} {\bibfnamefont {C.}~\bibnamefont {Stock}}, \ and\ \bibinfo
  {author} {\bibfnamefont {J.}~\bibnamefont {Taylor}},\ }\href {\doibase
  10.1103/PhysRevLett.108.127204} {\bibfield  {journal} {\bibinfo  {journal}
  {Phys. Rev. Lett.}\ }\textbf {\bibinfo {volume} {108}},\ \bibinfo {pages}
  {127204} (\bibinfo {year} {2012})}\BibitemShut {NoStop}%
\bibitem [{\citenamefont {Kim}\ \emph {et~al.}(2012{\natexlab{b}})\citenamefont
  {Kim}, \citenamefont {Said}, \citenamefont {Casa}, \citenamefont {Upton},
  \citenamefont {Gog}, \citenamefont {Daghofer}, \citenamefont {Jackeli},
  \citenamefont {van~den Brink}, \citenamefont {Khaliullin},\ and\
  \citenamefont {Kim}}]{JKim2012b}%
  \BibitemOpen
  \bibfield  {author} {\bibinfo {author} {\bibfnamefont {J.}~\bibnamefont
  {Kim}}, \bibinfo {author} {\bibfnamefont {A.~H.}\ \bibnamefont {Said}},
  \bibinfo {author} {\bibfnamefont {D.}~\bibnamefont {Casa}}, \bibinfo {author}
  {\bibfnamefont {M.~H.}\ \bibnamefont {Upton}}, \bibinfo {author}
  {\bibfnamefont {T.}~\bibnamefont {Gog}}, \bibinfo {author} {\bibfnamefont
  {M.}~\bibnamefont {Daghofer}}, \bibinfo {author} {\bibfnamefont
  {G.}~\bibnamefont {Jackeli}}, \bibinfo {author} {\bibfnamefont
  {J.}~\bibnamefont {van~den Brink}}, \bibinfo {author} {\bibfnamefont
  {G.}~\bibnamefont {Khaliullin}}, \ and\ \bibinfo {author} {\bibfnamefont
  {B.~J.}\ \bibnamefont {Kim}},\ }\href {\doibase
  10.1103/PhysRevLett.109.157402} {\bibfield  {journal} {\bibinfo  {journal}
  {Phys. Rev. Lett.}\ }\textbf {\bibinfo {volume} {109}},\ \bibinfo {pages}
  {157402} (\bibinfo {year} {2012}{\natexlab{b}})}\BibitemShut {NoStop}%
\bibitem [{\citenamefont {Moretti~Sala}\ \emph {et~al.}(2015)\citenamefont
  {Moretti~Sala}, \citenamefont {Schnells}, \citenamefont {Boseggia},
  \citenamefont {Simonelli}, \citenamefont {Al-Zein}, \citenamefont {Vale},
  \citenamefont {Paolasini}, \citenamefont {Hunter}, \citenamefont {Perry},
  \citenamefont {Prabhakaran}, \citenamefont {Boothroyd}, \citenamefont
  {Krisch}, \citenamefont {Monaco}, \citenamefont {R\o{}nnow}, \citenamefont
  {McMorrow},\ and\ \citenamefont {Mila}}]{MorettiSala2015}%
  \BibitemOpen
  \bibfield  {author} {\bibinfo {author} {\bibfnamefont {M.}~\bibnamefont
  {Moretti~Sala}}, \bibinfo {author} {\bibfnamefont {V.}~\bibnamefont
  {Schnells}}, \bibinfo {author} {\bibfnamefont {S.}~\bibnamefont {Boseggia}},
  \bibinfo {author} {\bibfnamefont {L.}~\bibnamefont {Simonelli}}, \bibinfo
  {author} {\bibfnamefont {A.}~\bibnamefont {Al-Zein}}, \bibinfo {author}
  {\bibfnamefont {J.~G.}\ \bibnamefont {Vale}}, \bibinfo {author}
  {\bibfnamefont {L.}~\bibnamefont {Paolasini}}, \bibinfo {author}
  {\bibfnamefont {E.~C.}\ \bibnamefont {Hunter}}, \bibinfo {author}
  {\bibfnamefont {R.~S.}\ \bibnamefont {Perry}}, \bibinfo {author}
  {\bibfnamefont {D.}~\bibnamefont {Prabhakaran}}, \bibinfo {author}
  {\bibfnamefont {A.~T.}\ \bibnamefont {Boothroyd}}, \bibinfo {author}
  {\bibfnamefont {M.}~\bibnamefont {Krisch}}, \bibinfo {author} {\bibfnamefont
  {G.}~\bibnamefont {Monaco}}, \bibinfo {author} {\bibfnamefont {H.~M.}\
  \bibnamefont {R\o{}nnow}}, \bibinfo {author} {\bibfnamefont {D.~F.}\
  \bibnamefont {McMorrow}}, \ and\ \bibinfo {author} {\bibfnamefont
  {F.}~\bibnamefont {Mila}},\ }\href {\doibase 10.1103/PhysRevB.92.024405}
  {\bibfield  {journal} {\bibinfo  {journal} {Phys. Rev. B}\ }\textbf {\bibinfo
  {volume} {92}},\ \bibinfo {pages} {024405} (\bibinfo {year}
  {2015})}\BibitemShut {NoStop}%
\bibitem [{\citenamefont {Hogan}\ \emph {et~al.}(2016)\citenamefont {Hogan},
  \citenamefont {Dally}, \citenamefont {Upton}, \citenamefont {Clancy},
  \citenamefont {Finkelstein}, \citenamefont {Kim}, \citenamefont {Graf},\ and\
  \citenamefont {Wilson}}]{Hogan2016}%
  \BibitemOpen
  \bibfield  {author} {\bibinfo {author} {\bibfnamefont {T.}~\bibnamefont
  {Hogan}}, \bibinfo {author} {\bibfnamefont {R.}~\bibnamefont {Dally}},
  \bibinfo {author} {\bibfnamefont {M.}~\bibnamefont {Upton}}, \bibinfo
  {author} {\bibfnamefont {J.~P.}\ \bibnamefont {Clancy}}, \bibinfo {author}
  {\bibfnamefont {K.}~\bibnamefont {Finkelstein}}, \bibinfo {author}
  {\bibfnamefont {Y.-J.}\ \bibnamefont {Kim}}, \bibinfo {author} {\bibfnamefont
  {M.~J.}\ \bibnamefont {Graf}}, \ and\ \bibinfo {author} {\bibfnamefont
  {S.~D.}\ \bibnamefont {Wilson}},\ }\href {\doibase
  10.1103/PhysRevB.94.100401} {\bibfield  {journal} {\bibinfo  {journal} {Phys.
  Rev. B}\ }\textbf {\bibinfo {volume} {94}},\ \bibinfo {pages} {100401}
  (\bibinfo {year} {2016})}\BibitemShut {NoStop}%
\bibitem [{\citenamefont {Lu}\ \emph {et~al.}(2017)\citenamefont {Lu},
  \citenamefont {McNally}, \citenamefont {Moretti~Sala}, \citenamefont
  {Terzic}, \citenamefont {Upton}, \citenamefont {Casa}, \citenamefont
  {Ingold}, \citenamefont {Cao},\ and\ \citenamefont {Schmitt}}]{Lu2017}%
  \BibitemOpen
  \bibfield  {author} {\bibinfo {author} {\bibfnamefont {X.}~\bibnamefont
  {Lu}}, \bibinfo {author} {\bibfnamefont {D.~E.}\ \bibnamefont {McNally}},
  \bibinfo {author} {\bibfnamefont {M.}~\bibnamefont {Moretti~Sala}}, \bibinfo
  {author} {\bibfnamefont {J.}~\bibnamefont {Terzic}}, \bibinfo {author}
  {\bibfnamefont {M.~H.}\ \bibnamefont {Upton}}, \bibinfo {author}
  {\bibfnamefont {D.}~\bibnamefont {Casa}}, \bibinfo {author} {\bibfnamefont
  {G.}~\bibnamefont {Ingold}}, \bibinfo {author} {\bibfnamefont
  {G.}~\bibnamefont {Cao}}, \ and\ \bibinfo {author} {\bibfnamefont
  {T.}~\bibnamefont {Schmitt}},\ }\href {\doibase
  10.1103/PhysRevLett.118.027202} {\bibfield  {journal} {\bibinfo  {journal}
  {Phys. Rev. Lett.}\ }\textbf {\bibinfo {volume} {118}},\ \bibinfo {pages}
  {027202} (\bibinfo {year} {2017})}\BibitemShut {NoStop}%
\bibitem [{\citenamefont {Lupascu}\ \emph {et~al.}(2014)\citenamefont
  {Lupascu}, \citenamefont {Clancy}, \citenamefont {Gretarsson}, \citenamefont
  {Nie}, \citenamefont {Nichols}, \citenamefont {Terzic}, \citenamefont {Cao},
  \citenamefont {Seo}, \citenamefont {Islam}, \citenamefont {Upton},
  \citenamefont {Kim}, \citenamefont {Casa}, \citenamefont {Gog}, \citenamefont
  {Said}, \citenamefont {Katukuri}, \citenamefont {Stoll}, \citenamefont
  {Hozoi}, \citenamefont {van~den Brink},\ and\ \citenamefont
  {Kim}}]{Lupascu2014}%
  \BibitemOpen
  \bibfield  {author} {\bibinfo {author} {\bibfnamefont {A.}~\bibnamefont
  {Lupascu}}, \bibinfo {author} {\bibfnamefont {J.~P.}\ \bibnamefont {Clancy}},
  \bibinfo {author} {\bibfnamefont {H.}~\bibnamefont {Gretarsson}}, \bibinfo
  {author} {\bibfnamefont {Z.}~\bibnamefont {Nie}}, \bibinfo {author}
  {\bibfnamefont {J.}~\bibnamefont {Nichols}}, \bibinfo {author} {\bibfnamefont
  {J.}~\bibnamefont {Terzic}}, \bibinfo {author} {\bibfnamefont
  {G.}~\bibnamefont {Cao}}, \bibinfo {author} {\bibfnamefont {S.~S.~A.}\
  \bibnamefont {Seo}}, \bibinfo {author} {\bibfnamefont {Z.}~\bibnamefont
  {Islam}}, \bibinfo {author} {\bibfnamefont {M.~H.}\ \bibnamefont {Upton}},
  \bibinfo {author} {\bibfnamefont {J.}~\bibnamefont {Kim}}, \bibinfo {author}
  {\bibfnamefont {D.}~\bibnamefont {Casa}}, \bibinfo {author} {\bibfnamefont
  {T.}~\bibnamefont {Gog}}, \bibinfo {author} {\bibfnamefont {A.~H.}\
  \bibnamefont {Said}}, \bibinfo {author} {\bibfnamefont {V.~M.}\ \bibnamefont
  {Katukuri}}, \bibinfo {author} {\bibfnamefont {H.}~\bibnamefont {Stoll}},
  \bibinfo {author} {\bibfnamefont {L.}~\bibnamefont {Hozoi}}, \bibinfo
  {author} {\bibfnamefont {J.}~\bibnamefont {van~den Brink}}, \ and\ \bibinfo
  {author} {\bibfnamefont {Y.-J.}\ \bibnamefont {Kim}},\ }\href {\doibase
  10.1103/PhysRevLett.112.147201} {\bibfield  {journal} {\bibinfo  {journal}
  {Physical Review Letters}\ }\textbf {\bibinfo {volume} {112}} (\bibinfo
  {year} {2014}),\ 10.1103/PhysRevLett.112.147201}\BibitemShut {NoStop}%
\bibitem [{\citenamefont {Liu}\ \emph {et~al.}(2016)\citenamefont {Liu},
  \citenamefont {Dean}, \citenamefont {Meng}, \citenamefont {Upton},
  \citenamefont {Qi}, \citenamefont {Gog}, \citenamefont {Cao}, \citenamefont
  {Lin}, \citenamefont {Meyers}, \citenamefont {Ding}, \citenamefont {Cao},\
  and\ \citenamefont {Hill}}]{Liu2016}%
  \BibitemOpen
  \bibfield  {author} {\bibinfo {author} {\bibfnamefont {X.}~\bibnamefont
  {Liu}}, \bibinfo {author} {\bibfnamefont {M.~P.~M.}\ \bibnamefont {Dean}},
  \bibinfo {author} {\bibfnamefont {Z.~Y.}\ \bibnamefont {Meng}}, \bibinfo
  {author} {\bibfnamefont {M.~H.}\ \bibnamefont {Upton}}, \bibinfo {author}
  {\bibfnamefont {T.}~\bibnamefont {Qi}}, \bibinfo {author} {\bibfnamefont
  {T.}~\bibnamefont {Gog}}, \bibinfo {author} {\bibfnamefont {Y.}~\bibnamefont
  {Cao}}, \bibinfo {author} {\bibfnamefont {J.~Q.}\ \bibnamefont {Lin}},
  \bibinfo {author} {\bibfnamefont {D.}~\bibnamefont {Meyers}}, \bibinfo
  {author} {\bibfnamefont {H.}~\bibnamefont {Ding}}, \bibinfo {author}
  {\bibfnamefont {G.}~\bibnamefont {Cao}}, \ and\ \bibinfo {author}
  {\bibfnamefont {J.~P.}\ \bibnamefont {Hill}},\ }\href {\doibase
  10.1103/PhysRevB.93.241102} {\bibfield  {journal} {\bibinfo  {journal} {Phys.
  Rev. B}\ }\textbf {\bibinfo {volume} {93}},\ \bibinfo {pages} {241102}
  (\bibinfo {year} {2016})}\BibitemShut {NoStop}%
\bibitem [{\citenamefont {Gretarsson}\ \emph {et~al.}(2016)\citenamefont
  {Gretarsson}, \citenamefont {Sung}, \citenamefont {Porras}, \citenamefont
  {Bertinshaw}, \citenamefont {Dietl}, \citenamefont {Bruin}, \citenamefont
  {Bangura}, \citenamefont {Kim}, \citenamefont {Dinnebier}, \citenamefont
  {Kim}, \citenamefont {Al-Zein}, \citenamefont {Moretti~Sala}, \citenamefont
  {Krisch}, \citenamefont {Le~Tacon}, \citenamefont {Keimer},\ and\
  \citenamefont {Kim}}]{Gretarsson2016}%
  \BibitemOpen
  \bibfield  {author} {\bibinfo {author} {\bibfnamefont {H.}~\bibnamefont
  {Gretarsson}}, \bibinfo {author} {\bibfnamefont {N.~H.}\ \bibnamefont
  {Sung}}, \bibinfo {author} {\bibfnamefont {J.}~\bibnamefont {Porras}},
  \bibinfo {author} {\bibfnamefont {J.}~\bibnamefont {Bertinshaw}}, \bibinfo
  {author} {\bibfnamefont {C.}~\bibnamefont {Dietl}}, \bibinfo {author}
  {\bibfnamefont {J.~A.~N.}\ \bibnamefont {Bruin}}, \bibinfo {author}
  {\bibfnamefont {A.~F.}\ \bibnamefont {Bangura}}, \bibinfo {author}
  {\bibfnamefont {Y.~K.}\ \bibnamefont {Kim}}, \bibinfo {author} {\bibfnamefont
  {R.}~\bibnamefont {Dinnebier}}, \bibinfo {author} {\bibfnamefont
  {J.}~\bibnamefont {Kim}}, \bibinfo {author} {\bibfnamefont {A.}~\bibnamefont
  {Al-Zein}}, \bibinfo {author} {\bibfnamefont {M.}~\bibnamefont
  {Moretti~Sala}}, \bibinfo {author} {\bibfnamefont {M.}~\bibnamefont
  {Krisch}}, \bibinfo {author} {\bibfnamefont {M.}~\bibnamefont {Le~Tacon}},
  \bibinfo {author} {\bibfnamefont {B.}~\bibnamefont {Keimer}}, \ and\ \bibinfo
  {author} {\bibfnamefont {B.~J.}\ \bibnamefont {Kim}},\ }\href {\doibase
  10.1103/PhysRevLett.117.107001} {\bibfield  {journal} {\bibinfo  {journal}
  {Phys. Rev. Lett.}\ }\textbf {\bibinfo {volume} {117}},\ \bibinfo {pages}
  {107001} (\bibinfo {year} {2016})}\BibitemShut {NoStop}%
\bibitem [{\citenamefont {Pincini}\ \emph {et~al.}(2017)\citenamefont
  {Pincini}, \citenamefont {Vale}, \citenamefont {Donnerer}, \citenamefont
  {de~la Torre}, \citenamefont {Hunter}, \citenamefont {Perry}, \citenamefont
  {Moretti~Sala}, \citenamefont {Baumberger},\ and\ \citenamefont
  {McMorrow}}]{Pincini2017}%
  \BibitemOpen
  \bibfield  {author} {\bibinfo {author} {\bibfnamefont {D.}~\bibnamefont
  {Pincini}}, \bibinfo {author} {\bibfnamefont {J.~G.}\ \bibnamefont {Vale}},
  \bibinfo {author} {\bibfnamefont {C.}~\bibnamefont {Donnerer}}, \bibinfo
  {author} {\bibfnamefont {A.}~\bibnamefont {de~la Torre}}, \bibinfo {author}
  {\bibfnamefont {E.~C.}\ \bibnamefont {Hunter}}, \bibinfo {author}
  {\bibfnamefont {R.}~\bibnamefont {Perry}}, \bibinfo {author} {\bibfnamefont
  {M.}~\bibnamefont {Moretti~Sala}}, \bibinfo {author} {\bibfnamefont
  {F.}~\bibnamefont {Baumberger}}, \ and\ \bibinfo {author} {\bibfnamefont
  {D.~F.}\ \bibnamefont {McMorrow}},\ }\href {\doibase
  10.1103/PhysRevB.96.075162} {\bibfield  {journal} {\bibinfo  {journal} {Phys.
  Rev. B}\ }\textbf {\bibinfo {volume} {96}},\ \bibinfo {pages} {075162}
  (\bibinfo {year} {2017})}\BibitemShut {NoStop}%
\bibitem [{\citenamefont {Cao}\ \emph {et~al.}(2017)\citenamefont {Cao},
  \citenamefont {Liu}, \citenamefont {Xu}, \citenamefont {Yin}, \citenamefont
  {Meyers}, \citenamefont {Kim}, \citenamefont {Casa}, \citenamefont {Upton},
  \citenamefont {Gog}, \citenamefont {Berlijn}, \citenamefont {Alvarez},
  \citenamefont {Yuan}, \citenamefont {Terzic}, \citenamefont {Tranquada},
  \citenamefont {Hill}, \citenamefont {Cao}, \citenamefont {Konik},\ and\
  \citenamefont {Dean}}]{Cao2017}%
  \BibitemOpen
  \bibfield  {author} {\bibinfo {author} {\bibfnamefont {Y.}~\bibnamefont
  {Cao}}, \bibinfo {author} {\bibfnamefont {X.}~\bibnamefont {Liu}}, \bibinfo
  {author} {\bibfnamefont {W.}~\bibnamefont {Xu}}, \bibinfo {author}
  {\bibfnamefont {W.-G.}\ \bibnamefont {Yin}}, \bibinfo {author} {\bibfnamefont
  {D.}~\bibnamefont {Meyers}}, \bibinfo {author} {\bibfnamefont
  {J.}~\bibnamefont {Kim}}, \bibinfo {author} {\bibfnamefont {D.}~\bibnamefont
  {Casa}}, \bibinfo {author} {\bibfnamefont {M.~H.}\ \bibnamefont {Upton}},
  \bibinfo {author} {\bibfnamefont {T.}~\bibnamefont {Gog}}, \bibinfo {author}
  {\bibfnamefont {T.}~\bibnamefont {Berlijn}}, \bibinfo {author} {\bibfnamefont
  {G.}~\bibnamefont {Alvarez}}, \bibinfo {author} {\bibfnamefont
  {S.}~\bibnamefont {Yuan}}, \bibinfo {author} {\bibfnamefont {J.}~\bibnamefont
  {Terzic}}, \bibinfo {author} {\bibfnamefont {J.~M.}\ \bibnamefont
  {Tranquada}}, \bibinfo {author} {\bibfnamefont {J.~P.}\ \bibnamefont {Hill}},
  \bibinfo {author} {\bibfnamefont {G.}~\bibnamefont {Cao}}, \bibinfo {author}
  {\bibfnamefont {R.~M.}\ \bibnamefont {Konik}}, \ and\ \bibinfo {author}
  {\bibfnamefont {M.~P.~M.}\ \bibnamefont {Dean}},\ }\href {\doibase
  10.1103/PhysRevB.95.121103} {\bibfield  {journal} {\bibinfo  {journal} {Phys.
  Rev. B}\ }\textbf {\bibinfo {volume} {95}},\ \bibinfo {pages} {121103}
  (\bibinfo {year} {2017})}\BibitemShut {NoStop}%
\bibitem [{\citenamefont {Meyers}\ \emph {et~al.}(2017)\citenamefont {Meyers},
  \citenamefont {Cao}, \citenamefont {Fabbris}, \citenamefont {Robinson},
  \citenamefont {Hao}, \citenamefont {Frederick}, \citenamefont {Traynor},
  \citenamefont {Yang}, \citenamefont {Lin}, \citenamefont {Upton},
  \citenamefont {Casa}, \citenamefont {Kim}, \citenamefont {Gog}, \citenamefont
  {Karapetrova}, \citenamefont {Choi}, \citenamefont {Haskel}, \citenamefont
  {Ryan}, \citenamefont {Horak}, \citenamefont {Liu}, \citenamefont {Liu},\
  and\ \citenamefont {Dean}}]{Meyers2017}%
  \BibitemOpen
  \bibfield  {author} {\bibinfo {author} {\bibfnamefont {D.}~\bibnamefont
  {Meyers}}, \bibinfo {author} {\bibfnamefont {Y.}~\bibnamefont {Cao}},
  \bibinfo {author} {\bibfnamefont {G.}~\bibnamefont {Fabbris}}, \bibinfo
  {author} {\bibfnamefont {N.~J.}\ \bibnamefont {Robinson}}, \bibinfo {author}
  {\bibfnamefont {L.}~\bibnamefont {Hao}}, \bibinfo {author} {\bibfnamefont
  {C.}~\bibnamefont {Frederick}}, \bibinfo {author} {\bibfnamefont
  {N.}~\bibnamefont {Traynor}}, \bibinfo {author} {\bibfnamefont
  {J.}~\bibnamefont {Yang}}, \bibinfo {author} {\bibfnamefont {J.}~\bibnamefont
  {Lin}}, \bibinfo {author} {\bibfnamefont {M.~H.}\ \bibnamefont {Upton}},
  \bibinfo {author} {\bibfnamefont {D.}~\bibnamefont {Casa}}, \bibinfo {author}
  {\bibfnamefont {J.-W.}\ \bibnamefont {Kim}}, \bibinfo {author} {\bibfnamefont
  {T.}~\bibnamefont {Gog}}, \bibinfo {author} {\bibfnamefont {E.}~\bibnamefont
  {Karapetrova}}, \bibinfo {author} {\bibfnamefont {Y.}~\bibnamefont {Choi}},
  \bibinfo {author} {\bibfnamefont {D.}~\bibnamefont {Haskel}}, \bibinfo
  {author} {\bibfnamefont {P.~J.}\ \bibnamefont {Ryan}}, \bibinfo {author}
  {\bibfnamefont {L.}~\bibnamefont {Horak}}, \bibinfo {author} {\bibfnamefont
  {X.}~\bibnamefont {Liu}}, \bibinfo {author} {\bibfnamefont {J.}~\bibnamefont
  {Liu}}, \ and\ \bibinfo {author} {\bibfnamefont {M.~P.~M.}\ \bibnamefont
  {Dean}},\ }\href@noop {} {\bibfield  {journal} {\bibinfo  {journal} {e-print
  arXiv:1707.08910}\ } (\bibinfo {year} {2017})}\BibitemShut {NoStop}%
\bibitem [{\citenamefont {Gretarsson}\ \emph
  {et~al.}(2013{\natexlab{a}})\citenamefont {Gretarsson}, \citenamefont
  {Clancy}, \citenamefont {Singh}, \citenamefont {Gegenwart}, \citenamefont
  {Hill}, \citenamefont {Kim}, \citenamefont {Upton}, \citenamefont {Said},
  \citenamefont {Casa}, \citenamefont {Gog},\ and\ \citenamefont
  {Kim}}]{Gretarsson2013b}%
  \BibitemOpen
  \bibfield  {author} {\bibinfo {author} {\bibfnamefont {H.}~\bibnamefont
  {Gretarsson}}, \bibinfo {author} {\bibfnamefont {J.~P.}\ \bibnamefont
  {Clancy}}, \bibinfo {author} {\bibfnamefont {Y.}~\bibnamefont {Singh}},
  \bibinfo {author} {\bibfnamefont {P.}~\bibnamefont {Gegenwart}}, \bibinfo
  {author} {\bibfnamefont {J.~P.}\ \bibnamefont {Hill}}, \bibinfo {author}
  {\bibfnamefont {J.}~\bibnamefont {Kim}}, \bibinfo {author} {\bibfnamefont
  {M.~H.}\ \bibnamefont {Upton}}, \bibinfo {author} {\bibfnamefont {A.~H.}\
  \bibnamefont {Said}}, \bibinfo {author} {\bibfnamefont {D.}~\bibnamefont
  {Casa}}, \bibinfo {author} {\bibfnamefont {T.}~\bibnamefont {Gog}}, \ and\
  \bibinfo {author} {\bibfnamefont {Y.-J.}\ \bibnamefont {Kim}},\ }\href
  {\doibase 10.1103/PhysRevB.87.220407} {\bibfield  {journal} {\bibinfo
  {journal} {Phys. Rev. B}\ }\textbf {\bibinfo {volume} {87}},\ \bibinfo
  {pages} {220407} (\bibinfo {year} {2013}{\natexlab{a}})}\BibitemShut
  {NoStop}%
\bibitem [{\citenamefont {Donnerer}\ \emph {et~al.}(2016)\citenamefont
  {Donnerer}, \citenamefont {Rahn}, \citenamefont {Sala}, \citenamefont {Vale},
  \citenamefont {Pincini}, \citenamefont {Strempfer}, \citenamefont {Krisch},
  \citenamefont {Prabhakaran}, \citenamefont {Boothroyd},\ and\ \citenamefont
  {McMorrow}}]{Donnerer2016}%
  \BibitemOpen
  \bibfield  {author} {\bibinfo {author} {\bibfnamefont {C.}~\bibnamefont
  {Donnerer}}, \bibinfo {author} {\bibfnamefont {M.~C.}\ \bibnamefont {Rahn}},
  \bibinfo {author} {\bibfnamefont {M.~M.}\ \bibnamefont {Sala}}, \bibinfo
  {author} {\bibfnamefont {J.~G.}\ \bibnamefont {Vale}}, \bibinfo {author}
  {\bibfnamefont {D.}~\bibnamefont {Pincini}}, \bibinfo {author} {\bibfnamefont
  {J.}~\bibnamefont {Strempfer}}, \bibinfo {author} {\bibfnamefont
  {M.}~\bibnamefont {Krisch}}, \bibinfo {author} {\bibfnamefont
  {D.}~\bibnamefont {Prabhakaran}}, \bibinfo {author} {\bibfnamefont {A.~T.}\
  \bibnamefont {Boothroyd}}, \ and\ \bibinfo {author} {\bibfnamefont {D.~F.}\
  \bibnamefont {McMorrow}},\ }\href {\doibase 10.1103/PhysRevLett.117.037201}
  {\bibfield  {journal} {\bibinfo  {journal} {Phys. Rev. Lett.}\ }\textbf
  {\bibinfo {volume} {117}},\ \bibinfo {pages} {037201} (\bibinfo {year}
  {2016})}\BibitemShut {NoStop}%
\bibitem [{\citenamefont {Clancy}\ \emph {et~al.}(2016)\citenamefont {Clancy},
  \citenamefont {Gretarsson}, \citenamefont {Lee}, \citenamefont {Tian},
  \citenamefont {Kim}, \citenamefont {Upton}, \citenamefont {Casa},
  \citenamefont {Gog}, \citenamefont {Islam}, \citenamefont {Jeon},
  \citenamefont {Kim}, \citenamefont {Desgreniers}, \citenamefont {Kim},
  \citenamefont {Julian},\ and\ \citenamefont {Kim}}]{Clancy2016}%
  \BibitemOpen
  \bibfield  {author} {\bibinfo {author} {\bibfnamefont {J.~P.}\ \bibnamefont
  {Clancy}}, \bibinfo {author} {\bibfnamefont {H.}~\bibnamefont {Gretarsson}},
  \bibinfo {author} {\bibfnamefont {E.~K.~H.}\ \bibnamefont {Lee}}, \bibinfo
  {author} {\bibfnamefont {D.}~\bibnamefont {Tian}}, \bibinfo {author}
  {\bibfnamefont {J.}~\bibnamefont {Kim}}, \bibinfo {author} {\bibfnamefont
  {M.~H.}\ \bibnamefont {Upton}}, \bibinfo {author} {\bibfnamefont
  {D.}~\bibnamefont {Casa}}, \bibinfo {author} {\bibfnamefont {T.}~\bibnamefont
  {Gog}}, \bibinfo {author} {\bibfnamefont {Z.}~\bibnamefont {Islam}}, \bibinfo
  {author} {\bibfnamefont {B.-G.}\ \bibnamefont {Jeon}}, \bibinfo {author}
  {\bibfnamefont {K.~H.}\ \bibnamefont {Kim}}, \bibinfo {author} {\bibfnamefont
  {S.}~\bibnamefont {Desgreniers}}, \bibinfo {author} {\bibfnamefont {Y.~B.}\
  \bibnamefont {Kim}}, \bibinfo {author} {\bibfnamefont {S.~J.}\ \bibnamefont
  {Julian}}, \ and\ \bibinfo {author} {\bibfnamefont {Y.-J.}\ \bibnamefont
  {Kim}},\ }\href {\doibase 10.1103/PhysRevB.94.024408} {\bibfield  {journal}
  {\bibinfo  {journal} {Phys. Rev. B}\ }\textbf {\bibinfo {volume} {94}},\
  \bibinfo {pages} {024408} (\bibinfo {year} {2016})}\BibitemShut {NoStop}%
\bibitem [{\citenamefont {Chun}\ \emph {et~al.}(2018)\citenamefont {Chun},
  \citenamefont {Yuan}, \citenamefont {Casa}, \citenamefont {Kim},
  \citenamefont {Kim}, \citenamefont {Tian}, \citenamefont {Qiu}, \citenamefont
  {Nakatsuji},\ and\ \citenamefont {Kim}}]{Chun2018}%
  \BibitemOpen
  \bibfield  {author} {\bibinfo {author} {\bibfnamefont {S.~H.}\ \bibnamefont
  {Chun}}, \bibinfo {author} {\bibfnamefont {B.}~\bibnamefont {Yuan}}, \bibinfo
  {author} {\bibfnamefont {D.}~\bibnamefont {Casa}}, \bibinfo {author}
  {\bibfnamefont {J.}~\bibnamefont {Kim}}, \bibinfo {author} {\bibfnamefont
  {C.-Y.}\ \bibnamefont {Kim}}, \bibinfo {author} {\bibfnamefont
  {Z.}~\bibnamefont {Tian}}, \bibinfo {author} {\bibfnamefont {Y.}~\bibnamefont
  {Qiu}}, \bibinfo {author} {\bibfnamefont {S.}~\bibnamefont {Nakatsuji}}, \
  and\ \bibinfo {author} {\bibfnamefont {Y.-J.}\ \bibnamefont {Kim}},\ }\href
  {\doibase 10.1103/PhysRevLett.120.177203} {\bibfield  {journal} {\bibinfo
  {journal} {Phys. Rev. Lett.}\ }\textbf {\bibinfo {volume} {120}},\ \bibinfo
  {pages} {177203} (\bibinfo {year} {2018})}\BibitemShut {NoStop}%
\bibitem [{\citenamefont {Abragam}\ and\ \citenamefont
  {Pryce}(1951)}]{Abragam1951}%
  \BibitemOpen
  \bibfield  {author} {\bibinfo {author} {\bibfnamefont {A.}~\bibnamefont
  {Abragam}}\ and\ \bibinfo {author} {\bibfnamefont {M.~H.~L.}\ \bibnamefont
  {Pryce}},\ }\href {\doibase 10.1098/rspa.1951.0022} {\bibfield  {journal}
  {\bibinfo  {journal} {Proceedings of the Royal Society of London A:
  Mathematical, Physical and Engineering Sciences}\ }\textbf {\bibinfo {volume}
  {205}},\ \bibinfo {pages} {135} (\bibinfo {year} {1951})}\BibitemShut
  {NoStop}%
\bibitem [{\citenamefont {Hozoi}\ \emph {et~al.}(2014)\citenamefont {Hozoi},
  \citenamefont {Gretarsson}, \citenamefont {Clancy}, \citenamefont {Jeon},
  \citenamefont {Lee}, \citenamefont {Kim}, \citenamefont {Yushankhai},
  \citenamefont {Fulde}, \citenamefont {Casa}, \citenamefont {Gog},
  \citenamefont {Kim}, \citenamefont {Said}, \citenamefont {Upton},
  \citenamefont {Kim},\ and\ \citenamefont {van~den Brink}}]{Hozoi2014}%
  \BibitemOpen
  \bibfield  {author} {\bibinfo {author} {\bibfnamefont {L.}~\bibnamefont
  {Hozoi}}, \bibinfo {author} {\bibfnamefont {H.}~\bibnamefont {Gretarsson}},
  \bibinfo {author} {\bibfnamefont {J.~P.}\ \bibnamefont {Clancy}}, \bibinfo
  {author} {\bibfnamefont {B.-G.}\ \bibnamefont {Jeon}}, \bibinfo {author}
  {\bibfnamefont {B.}~\bibnamefont {Lee}}, \bibinfo {author} {\bibfnamefont
  {K.~H.}\ \bibnamefont {Kim}}, \bibinfo {author} {\bibfnamefont
  {V.}~\bibnamefont {Yushankhai}}, \bibinfo {author} {\bibfnamefont
  {P.}~\bibnamefont {Fulde}}, \bibinfo {author} {\bibfnamefont
  {D.}~\bibnamefont {Casa}}, \bibinfo {author} {\bibfnamefont {T.}~\bibnamefont
  {Gog}}, \bibinfo {author} {\bibfnamefont {J.}~\bibnamefont {Kim}}, \bibinfo
  {author} {\bibfnamefont {A.~H.}\ \bibnamefont {Said}}, \bibinfo {author}
  {\bibfnamefont {M.~H.}\ \bibnamefont {Upton}}, \bibinfo {author}
  {\bibfnamefont {Y.-J.}\ \bibnamefont {Kim}}, \ and\ \bibinfo {author}
  {\bibfnamefont {J.}~\bibnamefont {van~den Brink}},\ }\href {\doibase
  10.1103/PhysRevB.89.115111} {\bibfield  {journal} {\bibinfo  {journal} {Phys.
  Rev. B}\ }\textbf {\bibinfo {volume} {89}},\ \bibinfo {pages} {115111}
  (\bibinfo {year} {2014})}\BibitemShut {NoStop}%
\bibitem [{\citenamefont {Liu}\ \emph {et~al.}(2012)\citenamefont {Liu},
  \citenamefont {Katukuri}, \citenamefont {Hozoi}, \citenamefont {Yin},
  \citenamefont {Dean}, \citenamefont {Upton}, \citenamefont {Kim},
  \citenamefont {Casa}, \citenamefont {Said}, \citenamefont {Gog},
  \citenamefont {Qi}, \citenamefont {Cao}, \citenamefont {Tsvelik},
  \citenamefont {van~den Brink},\ and\ \citenamefont {Hill}}]{Liu2012}%
  \BibitemOpen
  \bibfield  {author} {\bibinfo {author} {\bibfnamefont {X.}~\bibnamefont
  {Liu}}, \bibinfo {author} {\bibfnamefont {V.~M.}\ \bibnamefont {Katukuri}},
  \bibinfo {author} {\bibfnamefont {L.}~\bibnamefont {Hozoi}}, \bibinfo
  {author} {\bibfnamefont {W.-G.}\ \bibnamefont {Yin}}, \bibinfo {author}
  {\bibfnamefont {M.~P.~M.}\ \bibnamefont {Dean}}, \bibinfo {author}
  {\bibfnamefont {M.~H.}\ \bibnamefont {Upton}}, \bibinfo {author}
  {\bibfnamefont {J.}~\bibnamefont {Kim}}, \bibinfo {author} {\bibfnamefont
  {D.}~\bibnamefont {Casa}}, \bibinfo {author} {\bibfnamefont {A.}~\bibnamefont
  {Said}}, \bibinfo {author} {\bibfnamefont {T.}~\bibnamefont {Gog}}, \bibinfo
  {author} {\bibfnamefont {T.~F.}\ \bibnamefont {Qi}}, \bibinfo {author}
  {\bibfnamefont {G.}~\bibnamefont {Cao}}, \bibinfo {author} {\bibfnamefont
  {A.~M.}\ \bibnamefont {Tsvelik}}, \bibinfo {author} {\bibfnamefont
  {J.}~\bibnamefont {van~den Brink}}, \ and\ \bibinfo {author} {\bibfnamefont
  {J.~P.}\ \bibnamefont {Hill}},\ }\href {\doibase
  10.1103/PhysRevLett.109.157401} {\bibfield  {journal} {\bibinfo  {journal}
  {Phys. Rev. Lett.}\ }\textbf {\bibinfo {volume} {109}},\ \bibinfo {pages}
  {157401} (\bibinfo {year} {2012})}\BibitemShut {NoStop}%
\bibitem [{\citenamefont {Gretarsson}\ \emph
  {et~al.}(2013{\natexlab{b}})\citenamefont {Gretarsson}, \citenamefont
  {Clancy}, \citenamefont {Liu}, \citenamefont {Hill}, \citenamefont {Bozin},
  \citenamefont {Singh}, \citenamefont {Manni}, \citenamefont {Gegenwart},
  \citenamefont {Kim}, \citenamefont {Said}, \citenamefont {Casa},
  \citenamefont {Gog}, \citenamefont {Upton}, \citenamefont {Kim},
  \citenamefont {Yu}, \citenamefont {Katukuri}, \citenamefont {Hozoi},
  \citenamefont {van~den Brink},\ and\ \citenamefont {Kim}}]{Gretarsson2013a}%
  \BibitemOpen
  \bibfield  {author} {\bibinfo {author} {\bibfnamefont {H.}~\bibnamefont
  {Gretarsson}}, \bibinfo {author} {\bibfnamefont {J.~P.}\ \bibnamefont
  {Clancy}}, \bibinfo {author} {\bibfnamefont {X.}~\bibnamefont {Liu}},
  \bibinfo {author} {\bibfnamefont {J.~P.}\ \bibnamefont {Hill}}, \bibinfo
  {author} {\bibfnamefont {E.}~\bibnamefont {Bozin}}, \bibinfo {author}
  {\bibfnamefont {Y.}~\bibnamefont {Singh}}, \bibinfo {author} {\bibfnamefont
  {S.}~\bibnamefont {Manni}}, \bibinfo {author} {\bibfnamefont
  {P.}~\bibnamefont {Gegenwart}}, \bibinfo {author} {\bibfnamefont
  {J.}~\bibnamefont {Kim}}, \bibinfo {author} {\bibfnamefont {A.~H.}\
  \bibnamefont {Said}}, \bibinfo {author} {\bibfnamefont {D.}~\bibnamefont
  {Casa}}, \bibinfo {author} {\bibfnamefont {T.}~\bibnamefont {Gog}}, \bibinfo
  {author} {\bibfnamefont {M.~H.}\ \bibnamefont {Upton}}, \bibinfo {author}
  {\bibfnamefont {H.-S.}\ \bibnamefont {Kim}}, \bibinfo {author} {\bibfnamefont
  {J.}~\bibnamefont {Yu}}, \bibinfo {author} {\bibfnamefont {V.~M.}\
  \bibnamefont {Katukuri}}, \bibinfo {author} {\bibfnamefont {L.}~\bibnamefont
  {Hozoi}}, \bibinfo {author} {\bibfnamefont {J.}~\bibnamefont {van~den
  Brink}}, \ and\ \bibinfo {author} {\bibfnamefont {Y.-J.}\ \bibnamefont
  {Kim}},\ }\href {\doibase 10.1103/PhysRevLett.110.076402} {\bibfield
  {journal} {\bibinfo  {journal} {Phys. Rev. Lett.}\ }\textbf {\bibinfo
  {volume} {110}},\ \bibinfo {pages} {076402} (\bibinfo {year}
  {2013}{\natexlab{b}})}\BibitemShut {NoStop}%
\bibitem [{\citenamefont {Moretti-Sala}\ \emph {et~al.}(2014)\citenamefont
  {Moretti-Sala}, \citenamefont {Ohgushi}, \citenamefont {Al-Zein},
  \citenamefont {Hirata}, \citenamefont {Monaco},\ and\ \citenamefont
  {Krisch}}]{MorettiSala2014a}%
  \BibitemOpen
  \bibfield  {author} {\bibinfo {author} {\bibfnamefont {M.}~\bibnamefont
  {Moretti-Sala}}, \bibinfo {author} {\bibfnamefont {K.}~\bibnamefont
  {Ohgushi}}, \bibinfo {author} {\bibfnamefont {A.}~\bibnamefont {Al-Zein}},
  \bibinfo {author} {\bibfnamefont {Y.}~\bibnamefont {Hirata}}, \bibinfo
  {author} {\bibfnamefont {G.}~\bibnamefont {Monaco}}, \ and\ \bibinfo {author}
  {\bibfnamefont {M.}~\bibnamefont {Krisch}},\ }\href {\doibase
  10.1103/PhysRevLett.112.176402} {\bibfield  {journal} {\bibinfo  {journal}
  {Phys. Rev. Lett.}\ }\textbf {\bibinfo {volume} {112}},\ \bibinfo {pages}
  {176402} (\bibinfo {year} {2014})}\BibitemShut {NoStop}%
\bibitem [{\citenamefont {Kim}\ \emph {et~al.}(2014)\citenamefont {Kim},
  \citenamefont {Daghofer}, \citenamefont {Said}, \citenamefont {Gog},
  \citenamefont {van~den Brink}, \citenamefont {Khaliullin},\ and\
  \citenamefont {Kim}}]{JKim2014}%
  \BibitemOpen
  \bibfield  {author} {\bibinfo {author} {\bibfnamefont {J.}~\bibnamefont
  {Kim}}, \bibinfo {author} {\bibfnamefont {M.}~\bibnamefont {Daghofer}},
  \bibinfo {author} {\bibfnamefont {A.~H.}\ \bibnamefont {Said}}, \bibinfo
  {author} {\bibfnamefont {T.}~\bibnamefont {Gog}}, \bibinfo {author}
  {\bibfnamefont {J.}~\bibnamefont {van~den Brink}}, \bibinfo {author}
  {\bibfnamefont {G.}~\bibnamefont {Khaliullin}}, \ and\ \bibinfo {author}
  {\bibfnamefont {B.~J.}\ \bibnamefont {Kim}},\ }\href@noop {} {\bibfield
  {journal} {\bibinfo  {journal} {Nat. Comm.}\ }\textbf {\bibinfo {volume}
  {5}},\ \bibinfo {pages} {4453} (\bibinfo {year} {2014})}\BibitemShut
  {NoStop}%
\bibitem [{\citenamefont {Bogdanov}\ \emph {et~al.}(2014)\citenamefont
  {Bogdanov}, \citenamefont {Katukuri}, \citenamefont {Romhanyi}, \citenamefont
  {Yushankhai}, \citenamefont {Kataev}, \citenamefont {Büchner}, \citenamefont
  {van~den Brink},\ and\ \citenamefont {Hozoi}}]{Bogdanov2014}%
  \BibitemOpen
  \bibfield  {author} {\bibinfo {author} {\bibfnamefont {N.~A.}\ \bibnamefont
  {Bogdanov}}, \bibinfo {author} {\bibfnamefont {V.~M.}\ \bibnamefont
  {Katukuri}}, \bibinfo {author} {\bibfnamefont {J.}~\bibnamefont {Romhanyi}},
  \bibinfo {author} {\bibfnamefont {V.}~\bibnamefont {Yushankhai}}, \bibinfo
  {author} {\bibfnamefont {V.}~\bibnamefont {Kataev}}, \bibinfo {author}
  {\bibfnamefont {B.}~\bibnamefont {Büchner}}, \bibinfo {author}
  {\bibfnamefont {J.}~\bibnamefont {van~den Brink}}, \ and\ \bibinfo {author}
  {\bibfnamefont {L.}~\bibnamefont {Hozoi}},\ }\href {\doibase
  doi:10.1038/ncomms8306} {\bibfield  {journal} {\bibinfo  {journal} {Nat.
  Comm.}\ }\textbf {\bibinfo {volume} {6}},\ \bibinfo {pages} {7306} (\bibinfo
  {year} {2014})}\BibitemShut {NoStop}%
\bibitem [{\citenamefont {Yuan}\ \emph {et~al.}(2017)\citenamefont {Yuan},
  \citenamefont {Clancy}, \citenamefont {Cook}, \citenamefont {Thompson},
  \citenamefont {Greedan}, \citenamefont {Cao}, \citenamefont {Jeon},
  \citenamefont {Noh}, \citenamefont {Upton}, \citenamefont {Casa},
  \citenamefont {Gog}, \citenamefont {Paramekanti},\ and\ \citenamefont
  {Kim}}]{Yuan2017}%
  \BibitemOpen
  \bibfield  {author} {\bibinfo {author} {\bibfnamefont {B.}~\bibnamefont
  {Yuan}}, \bibinfo {author} {\bibfnamefont {J.~P.}\ \bibnamefont {Clancy}},
  \bibinfo {author} {\bibfnamefont {A.~M.}\ \bibnamefont {Cook}}, \bibinfo
  {author} {\bibfnamefont {C.~M.}\ \bibnamefont {Thompson}}, \bibinfo {author}
  {\bibfnamefont {J.}~\bibnamefont {Greedan}}, \bibinfo {author} {\bibfnamefont
  {G.}~\bibnamefont {Cao}}, \bibinfo {author} {\bibfnamefont {B.~C.}\
  \bibnamefont {Jeon}}, \bibinfo {author} {\bibfnamefont {T.~W.}\ \bibnamefont
  {Noh}}, \bibinfo {author} {\bibfnamefont {M.~H.}\ \bibnamefont {Upton}},
  \bibinfo {author} {\bibfnamefont {D.}~\bibnamefont {Casa}}, \bibinfo {author}
  {\bibfnamefont {T.}~\bibnamefont {Gog}}, \bibinfo {author} {\bibfnamefont
  {A.}~\bibnamefont {Paramekanti}}, \ and\ \bibinfo {author} {\bibfnamefont
  {Y.-J.}\ \bibnamefont {Kim}},\ }\href {\doibase 10.1103/PhysRevB.95.235114}
  {\bibfield  {journal} {\bibinfo  {journal} {Phys. Rev. B}\ }\textbf {\bibinfo
  {volume} {95}},\ \bibinfo {pages} {235114} (\bibinfo {year}
  {2017})}\BibitemShut {NoStop}%
\bibitem [{\citenamefont {Lu}\ \emph {et~al.}(2006)\citenamefont {Lu},
  \citenamefont {Hancock}, \citenamefont {Chabot-Couture}, \citenamefont
  {Ishii}, \citenamefont {Vajk}, \citenamefont {Yu}, \citenamefont {Mizuki},
  \citenamefont {Casa}, \citenamefont {Gog},\ and\ \citenamefont
  {Greven}}]{Lu2006}%
  \BibitemOpen
  \bibfield  {author} {\bibinfo {author} {\bibfnamefont {L.}~\bibnamefont
  {Lu}}, \bibinfo {author} {\bibfnamefont {J.~N.}\ \bibnamefont {Hancock}},
  \bibinfo {author} {\bibfnamefont {G.}~\bibnamefont {Chabot-Couture}},
  \bibinfo {author} {\bibfnamefont {K.}~\bibnamefont {Ishii}}, \bibinfo
  {author} {\bibfnamefont {O.~P.}\ \bibnamefont {Vajk}}, \bibinfo {author}
  {\bibfnamefont {G.}~\bibnamefont {Yu}}, \bibinfo {author} {\bibfnamefont
  {J.}~\bibnamefont {Mizuki}}, \bibinfo {author} {\bibfnamefont
  {D.}~\bibnamefont {Casa}}, \bibinfo {author} {\bibfnamefont {T.}~\bibnamefont
  {Gog}}, \ and\ \bibinfo {author} {\bibfnamefont {M.}~\bibnamefont {Greven}},\
  }\href {\doibase 10.1103/PhysRevB.74.224509} {\bibfield  {journal} {\bibinfo
  {journal} {Phys. Rev. B}\ }\textbf {\bibinfo {volume} {74}},\ \bibinfo
  {pages} {224509} (\bibinfo {year} {2006})}\BibitemShut {NoStop}%
\bibitem [{\citenamefont {Ghiringhelli}\ \emph {et~al.}(2009)\citenamefont
  {Ghiringhelli}, \citenamefont {Piazzalunga}, \citenamefont {Dallera},
  \citenamefont {Schmitt}, \citenamefont {Strocov}, \citenamefont {Schlappa},
  \citenamefont {Patthey}, \citenamefont {Wang}, \citenamefont {Berger},\ and\
  \citenamefont {Grioni}}]{Ghiringhelli2009}%
  \BibitemOpen
  \bibfield  {author} {\bibinfo {author} {\bibfnamefont {G.}~\bibnamefont
  {Ghiringhelli}}, \bibinfo {author} {\bibfnamefont {A.}~\bibnamefont
  {Piazzalunga}}, \bibinfo {author} {\bibfnamefont {C.}~\bibnamefont
  {Dallera}}, \bibinfo {author} {\bibfnamefont {T.}~\bibnamefont {Schmitt}},
  \bibinfo {author} {\bibfnamefont {V.~N.}\ \bibnamefont {Strocov}}, \bibinfo
  {author} {\bibfnamefont {J.}~\bibnamefont {Schlappa}}, \bibinfo {author}
  {\bibfnamefont {L.}~\bibnamefont {Patthey}}, \bibinfo {author} {\bibfnamefont
  {X.}~\bibnamefont {Wang}}, \bibinfo {author} {\bibfnamefont {H.}~\bibnamefont
  {Berger}}, \ and\ \bibinfo {author} {\bibfnamefont {M.}~\bibnamefont
  {Grioni}},\ }\href {\doibase 10.1103/PhysRevLett.102.027401} {\bibfield
  {journal} {\bibinfo  {journal} {Phys. Rev. Lett.}\ }\textbf {\bibinfo
  {volume} {102}},\ \bibinfo {pages} {027401} (\bibinfo {year}
  {2009})}\BibitemShut {NoStop}%
\bibitem [{\citenamefont {Harada}\ \emph {et~al.}(2000)\citenamefont {Harada},
  \citenamefont {Kinugasa}, \citenamefont {Eguchi}, \citenamefont {Matsubara},
  \citenamefont {Kotani}, \citenamefont {Watanabe}, \citenamefont {Yagishita},\
  and\ \citenamefont {Shin}}]{Harada2000}%
  \BibitemOpen
  \bibfield  {author} {\bibinfo {author} {\bibfnamefont {Y.}~\bibnamefont
  {Harada}}, \bibinfo {author} {\bibfnamefont {T.}~\bibnamefont {Kinugasa}},
  \bibinfo {author} {\bibfnamefont {R.}~\bibnamefont {Eguchi}}, \bibinfo
  {author} {\bibfnamefont {M.}~\bibnamefont {Matsubara}}, \bibinfo {author}
  {\bibfnamefont {A.}~\bibnamefont {Kotani}}, \bibinfo {author} {\bibfnamefont
  {M.}~\bibnamefont {Watanabe}}, \bibinfo {author} {\bibfnamefont
  {A.}~\bibnamefont {Yagishita}}, \ and\ \bibinfo {author} {\bibfnamefont
  {S.}~\bibnamefont {Shin}},\ }\href {\doibase 10.1103/PhysRevB.61.12854}
  {\bibfield  {journal} {\bibinfo  {journal} {Phys. Rev. B}\ }\textbf {\bibinfo
  {volume} {61}},\ \bibinfo {pages} {12854} (\bibinfo {year}
  {2000})}\BibitemShut {NoStop}%
\bibitem [{\citenamefont {Gretarsson}\ \emph {et~al.}(2011)\citenamefont
  {Gretarsson}, \citenamefont {Kim}, \citenamefont {Casa}, \citenamefont {Gog},
  \citenamefont {Choi}, \citenamefont {Cheong},\ and\ \citenamefont
  {Kim}}]{Gretarsson2011}%
  \BibitemOpen
  \bibfield  {author} {\bibinfo {author} {\bibfnamefont {H.}~\bibnamefont
  {Gretarsson}}, \bibinfo {author} {\bibfnamefont {J.}~\bibnamefont {Kim}},
  \bibinfo {author} {\bibfnamefont {D.}~\bibnamefont {Casa}}, \bibinfo {author}
  {\bibfnamefont {T.}~\bibnamefont {Gog}}, \bibinfo {author} {\bibfnamefont
  {K.~R.}\ \bibnamefont {Choi}}, \bibinfo {author} {\bibfnamefont {S.~W.}\
  \bibnamefont {Cheong}}, \ and\ \bibinfo {author} {\bibfnamefont {Y.-J.}\
  \bibnamefont {Kim}},\ }\href {\doibase 10.1103/PhysRevB.84.125135} {\bibfield
   {journal} {\bibinfo  {journal} {Phys. Rev. B}\ }\textbf {\bibinfo {volume}
  {84}},\ \bibinfo {pages} {125135} (\bibinfo {year} {2011})}\BibitemShut
  {NoStop}%
\bibitem [{\citenamefont {Clancy}\ \emph {et~al.}(2014)\citenamefont {Clancy},
  \citenamefont {Lupascu}, \citenamefont {Gretarsson}, \citenamefont {Islam},
  \citenamefont {Hu}, \citenamefont {Casa}, \citenamefont {Nelson},
  \citenamefont {LaMarra}, \citenamefont {Cao},\ and\ \citenamefont
  {Kim}}]{Clancy2014}%
  \BibitemOpen
  \bibfield  {author} {\bibinfo {author} {\bibfnamefont {J.~P.}\ \bibnamefont
  {Clancy}}, \bibinfo {author} {\bibfnamefont {A.}~\bibnamefont {Lupascu}},
  \bibinfo {author} {\bibfnamefont {H.}~\bibnamefont {Gretarsson}}, \bibinfo
  {author} {\bibfnamefont {Z.}~\bibnamefont {Islam}}, \bibinfo {author}
  {\bibfnamefont {Y.~F.}\ \bibnamefont {Hu}}, \bibinfo {author} {\bibfnamefont
  {D.}~\bibnamefont {Casa}}, \bibinfo {author} {\bibfnamefont {C.~S.}\
  \bibnamefont {Nelson}}, \bibinfo {author} {\bibfnamefont {S.~C.}\
  \bibnamefont {LaMarra}}, \bibinfo {author} {\bibfnamefont {G.}~\bibnamefont
  {Cao}}, \ and\ \bibinfo {author} {\bibfnamefont {Y.-J.}\ \bibnamefont
  {Kim}},\ }\href {\doibase 10.1103/PhysRevB.89.054409} {\bibfield  {journal}
  {\bibinfo  {journal} {Phys. Rev. B}\ }\textbf {\bibinfo {volume} {89}},\
  \bibinfo {pages} {054409} (\bibinfo {year} {2014})}\BibitemShut {NoStop}%
\bibitem [{\citenamefont {Brodersen}\ \emph {et~al.}(1965)\citenamefont
  {Brodersen}, \citenamefont {Moers},\ and\ \citenamefont
  {Schnering}}]{Brodersen1965}%
  \BibitemOpen
  \bibfield  {author} {\bibinfo {author} {\bibfnamefont {K.}~\bibnamefont
  {Brodersen}}, \bibinfo {author} {\bibfnamefont {F.}~\bibnamefont {Moers}}, \
  and\ \bibinfo {author} {\bibfnamefont {H.}~\bibnamefont {Schnering}},\
  }\href@noop {} {\bibfield  {journal} {\bibinfo  {journal}
  {Naturwissenschaften}\ }\textbf {\bibinfo {volume} {52}},\ \bibinfo {pages}
  {205} (\bibinfo {year} {1965})}\BibitemShut {NoStop}%
\bibitem [{\citenamefont {Plumb}\ \emph {et~al.}(2014)\citenamefont {Plumb},
  \citenamefont {Clancy}, \citenamefont {Sandilands}, \citenamefont {Shankar},
  \citenamefont {Hu}, \citenamefont {Burch}, \citenamefont {Kee},\ and\
  \citenamefont {Kim}}]{Plumb2014}%
  \BibitemOpen
  \bibfield  {author} {\bibinfo {author} {\bibfnamefont {K.~W.}\ \bibnamefont
  {Plumb}}, \bibinfo {author} {\bibfnamefont {J.~P.}\ \bibnamefont {Clancy}},
  \bibinfo {author} {\bibfnamefont {L.~J.}\ \bibnamefont {Sandilands}},
  \bibinfo {author} {\bibfnamefont {V.~V.}\ \bibnamefont {Shankar}}, \bibinfo
  {author} {\bibfnamefont {Y.~F.}\ \bibnamefont {Hu}}, \bibinfo {author}
  {\bibfnamefont {K.~S.}\ \bibnamefont {Burch}}, \bibinfo {author}
  {\bibfnamefont {H.-Y.}\ \bibnamefont {Kee}}, \ and\ \bibinfo {author}
  {\bibfnamefont {Y.-J.}\ \bibnamefont {Kim}},\ }\href {\doibase
  10.1103/PhysRevB.90.041112} {\bibfield  {journal} {\bibinfo  {journal}
  {Physical Review B}\ }\textbf {\bibinfo {volume} {90}} (\bibinfo {year}
  {2014}),\ 10.1103/PhysRevB.90.041112}\BibitemShut {NoStop}%
\bibitem [{\citenamefont {Sears}\ \emph {et~al.}(2015)\citenamefont {Sears},
  \citenamefont {Songvilay}, \citenamefont {Plumb}, \citenamefont {Clancy},
  \citenamefont {Qiu}, \citenamefont {Zhao}, \citenamefont {Parshall},\ and\
  \citenamefont {Kim}}]{Sears2015}%
  \BibitemOpen
  \bibfield  {author} {\bibinfo {author} {\bibfnamefont {J.~A.}\ \bibnamefont
  {Sears}}, \bibinfo {author} {\bibfnamefont {M.}~\bibnamefont {Songvilay}},
  \bibinfo {author} {\bibfnamefont {K.~W.}\ \bibnamefont {Plumb}}, \bibinfo
  {author} {\bibfnamefont {J.~P.}\ \bibnamefont {Clancy}}, \bibinfo {author}
  {\bibfnamefont {Y.}~\bibnamefont {Qiu}}, \bibinfo {author} {\bibfnamefont
  {Y.}~\bibnamefont {Zhao}}, \bibinfo {author} {\bibfnamefont {D.}~\bibnamefont
  {Parshall}}, \ and\ \bibinfo {author} {\bibfnamefont {Y.-J.}\ \bibnamefont
  {Kim}},\ }\href {\doibase 10.1103/PhysRevB.91.144420} {\bibfield  {journal}
  {\bibinfo  {journal} {Physical Review B}\ }\textbf {\bibinfo {volume} {91}}
  (\bibinfo {year} {2015}),\ 10.1103/PhysRevB.91.144420}\BibitemShut {NoStop}%
\bibitem [{\citenamefont {Majumder}\ \emph {et~al.}(2015)\citenamefont
  {Majumder}, \citenamefont {Schmidt}, \citenamefont {Rosner}, \citenamefont
  {Tsirlin}, \citenamefont {Yasuoka},\ and\ \citenamefont
  {Baenitz}}]{Majumder2015}%
  \BibitemOpen
  \bibfield  {author} {\bibinfo {author} {\bibfnamefont {M.}~\bibnamefont
  {Majumder}}, \bibinfo {author} {\bibfnamefont {M.}~\bibnamefont {Schmidt}},
  \bibinfo {author} {\bibfnamefont {H.}~\bibnamefont {Rosner}}, \bibinfo
  {author} {\bibfnamefont {A.~A.}\ \bibnamefont {Tsirlin}}, \bibinfo {author}
  {\bibfnamefont {H.}~\bibnamefont {Yasuoka}}, \ and\ \bibinfo {author}
  {\bibfnamefont {M.}~\bibnamefont {Baenitz}},\ }\href {\doibase
  10.1103/PhysRevB.91.180401} {\bibfield  {journal} {\bibinfo  {journal} {Phys.
  Rev. B}\ }\textbf {\bibinfo {volume} {91}},\ \bibinfo {pages} {180401}
  (\bibinfo {year} {2015})}\BibitemShut {NoStop}%
\bibitem [{\citenamefont {Sandilands}\ \emph {et~al.}(2015)\citenamefont
  {Sandilands}, \citenamefont {Tian}, \citenamefont {Plumb}, \citenamefont
  {Kim},\ and\ \citenamefont {Burch}}]{Sandilands2015}%
  \BibitemOpen
  \bibfield  {author} {\bibinfo {author} {\bibfnamefont {L.~J.}\ \bibnamefont
  {Sandilands}}, \bibinfo {author} {\bibfnamefont {Y.}~\bibnamefont {Tian}},
  \bibinfo {author} {\bibfnamefont {K.~W.}\ \bibnamefont {Plumb}}, \bibinfo
  {author} {\bibfnamefont {Y.-J.}\ \bibnamefont {Kim}}, \ and\ \bibinfo
  {author} {\bibfnamefont {K.~S.}\ \bibnamefont {Burch}},\ }\href {\doibase
  10.1103/PhysRevLett.114.147201} {\bibfield  {journal} {\bibinfo  {journal}
  {Physical Review Letters}\ }\textbf {\bibinfo {volume} {114}} (\bibinfo
  {year} {2015}),\ 10.1103/PhysRevLett.114.147201}\BibitemShut {NoStop}%
\bibitem [{\citenamefont {Kubota}\ \emph {et~al.}(2015)\citenamefont {Kubota},
  \citenamefont {Tanaka}, \citenamefont {Ono}, \citenamefont {Narumi},\ and\
  \citenamefont {Kindo}}]{Kubota2015}%
  \BibitemOpen
  \bibfield  {author} {\bibinfo {author} {\bibfnamefont {Y.}~\bibnamefont
  {Kubota}}, \bibinfo {author} {\bibfnamefont {H.}~\bibnamefont {Tanaka}},
  \bibinfo {author} {\bibfnamefont {T.}~\bibnamefont {Ono}}, \bibinfo {author}
  {\bibfnamefont {Y.}~\bibnamefont {Narumi}}, \ and\ \bibinfo {author}
  {\bibfnamefont {K.}~\bibnamefont {Kindo}},\ }\href {\doibase
  10.1103/PhysRevB.91.094422} {\bibfield  {journal} {\bibinfo  {journal} {Phys.
  Rev. B}\ }\textbf {\bibinfo {volume} {91}},\ \bibinfo {pages} {094422}
  (\bibinfo {year} {2015})}\BibitemShut {NoStop}%
\bibitem [{\citenamefont {Kim}\ \emph {et~al.}(2015)\citenamefont {Kim},
  \citenamefont {V.}, \citenamefont {Catuneanu},\ and\ \citenamefont
  {Kee}}]{HSKim2015}%
  \BibitemOpen
  \bibfield  {author} {\bibinfo {author} {\bibfnamefont {H.-S.}\ \bibnamefont
  {Kim}}, \bibinfo {author} {\bibfnamefont {V.~S.}\ \bibnamefont {V.}},
  \bibinfo {author} {\bibfnamefont {A.}~\bibnamefont {Catuneanu}}, \ and\
  \bibinfo {author} {\bibfnamefont {H.-Y.}\ \bibnamefont {Kee}},\ }\href
  {\doibase 10.1103/PhysRevB.91.241110} {\bibfield  {journal} {\bibinfo
  {journal} {Phys. Rev. B}\ }\textbf {\bibinfo {volume} {91}},\ \bibinfo
  {pages} {241110} (\bibinfo {year} {2015})}\BibitemShut {NoStop}%
\bibitem [{\citenamefont {Banerjee}\ \emph {et~al.}(2016)\citenamefont
  {Banerjee}, \citenamefont {Bridges}, \citenamefont {Yan}, \citenamefont
  {Aczel}, \citenamefont {Li}, \citenamefont {Stone}, \citenamefont {Granroth},
  \citenamefont {Lumsden}, \citenamefont {Yiu}, \citenamefont {Knolle},
  \citenamefont {Bhattacharjee}, \citenamefont {Kovrizhin}, \citenamefont
  {Moessner}, \citenamefont {Tennant}, \citenamefont {Mandrus},\ and\
  \citenamefont {Nagler}}]{Banerjee2016}%
  \BibitemOpen
  \bibfield  {author} {\bibinfo {author} {\bibfnamefont {A.}~\bibnamefont
  {Banerjee}}, \bibinfo {author} {\bibfnamefont {C.~A.}\ \bibnamefont
  {Bridges}}, \bibinfo {author} {\bibfnamefont {J.-Q.}\ \bibnamefont {Yan}},
  \bibinfo {author} {\bibfnamefont {A.~A.}\ \bibnamefont {Aczel}}, \bibinfo
  {author} {\bibfnamefont {L.}~\bibnamefont {Li}}, \bibinfo {author}
  {\bibfnamefont {M.~B.}\ \bibnamefont {Stone}}, \bibinfo {author}
  {\bibfnamefont {G.~E.}\ \bibnamefont {Granroth}}, \bibinfo {author}
  {\bibfnamefont {M.~D.}\ \bibnamefont {Lumsden}}, \bibinfo {author}
  {\bibfnamefont {Y.}~\bibnamefont {Yiu}}, \bibinfo {author} {\bibfnamefont
  {J.}~\bibnamefont {Knolle}}, \bibinfo {author} {\bibfnamefont
  {S.}~\bibnamefont {Bhattacharjee}}, \bibinfo {author} {\bibfnamefont {D.~L.}\
  \bibnamefont {Kovrizhin}}, \bibinfo {author} {\bibfnamefont {R.}~\bibnamefont
  {Moessner}}, \bibinfo {author} {\bibfnamefont {D.~A.}\ \bibnamefont
  {Tennant}}, \bibinfo {author} {\bibfnamefont {D.~G.}\ \bibnamefont
  {Mandrus}}, \ and\ \bibinfo {author} {\bibfnamefont {S.~E.}\ \bibnamefont
  {Nagler}},\ }\href@noop {} {\bibfield  {journal} {\bibinfo  {journal} {Nat.
  Mater.}\ } (\bibinfo {year} {2016})}\BibitemShut {NoStop}%
\bibitem [{\citenamefont {Cao}\ \emph {et~al.}(2014)\citenamefont {Cao},
  \citenamefont {Qi}, \citenamefont {Li}, \citenamefont {Terzic}, \citenamefont
  {Yuan}, \citenamefont {DeLong}, \citenamefont {Murthy},\ and\ \citenamefont
  {Kaul}}]{GCao2014}%
  \BibitemOpen
  \bibfield  {author} {\bibinfo {author} {\bibfnamefont {G.}~\bibnamefont
  {Cao}}, \bibinfo {author} {\bibfnamefont {T.~F.}\ \bibnamefont {Qi}},
  \bibinfo {author} {\bibfnamefont {L.}~\bibnamefont {Li}}, \bibinfo {author}
  {\bibfnamefont {J.}~\bibnamefont {Terzic}}, \bibinfo {author} {\bibfnamefont
  {S.~J.}\ \bibnamefont {Yuan}}, \bibinfo {author} {\bibfnamefont {L.~E.}\
  \bibnamefont {DeLong}}, \bibinfo {author} {\bibfnamefont {G.}~\bibnamefont
  {Murthy}}, \ and\ \bibinfo {author} {\bibfnamefont {R.~K.}\ \bibnamefont
  {Kaul}},\ }\href {\doibase 10.1103/PhysRevLett.112.056402} {\bibfield
  {journal} {\bibinfo  {journal} {Phys. Rev. Lett.}\ }\textbf {\bibinfo
  {volume} {112}},\ \bibinfo {pages} {056402} (\bibinfo {year}
  {2014})}\BibitemShut {NoStop}%
\bibitem [{\citenamefont {Qi}\ \emph {et~al.}(2012)\citenamefont {Qi},
  \citenamefont {Korneta}, \citenamefont {Li}, \citenamefont {Butrouna},
  \citenamefont {Cao}, \citenamefont {Wan}, \citenamefont {Schlottmann},
  \citenamefont {Kaul},\ and\ \citenamefont {Cao}}]{Qi2012}%
  \BibitemOpen
  \bibfield  {author} {\bibinfo {author} {\bibfnamefont {T.~F.}\ \bibnamefont
  {Qi}}, \bibinfo {author} {\bibfnamefont {O.~B.}\ \bibnamefont {Korneta}},
  \bibinfo {author} {\bibfnamefont {L.}~\bibnamefont {Li}}, \bibinfo {author}
  {\bibfnamefont {K.}~\bibnamefont {Butrouna}}, \bibinfo {author}
  {\bibfnamefont {V.~S.}\ \bibnamefont {Cao}}, \bibinfo {author} {\bibfnamefont
  {X.}~\bibnamefont {Wan}}, \bibinfo {author} {\bibfnamefont {P.}~\bibnamefont
  {Schlottmann}}, \bibinfo {author} {\bibfnamefont {R.~K.}\ \bibnamefont
  {Kaul}}, \ and\ \bibinfo {author} {\bibfnamefont {G.}~\bibnamefont {Cao}},\
  }\href {\doibase 10.1103/PhysRevB.86.125105} {\bibfield  {journal} {\bibinfo
  {journal} {Phys. Rev. B}\ }\textbf {\bibinfo {volume} {86}},\ \bibinfo
  {pages} {125105} (\bibinfo {year} {2012})}\BibitemShut {NoStop}%
\bibitem [{\citenamefont {Sarkozy}\ \emph {et~al.}(1974)\citenamefont
  {Sarkozy}, \citenamefont {Moeller},\ and\ \citenamefont
  {Chamberland}}]{Sarkozy1974}%
  \BibitemOpen
  \bibfield  {author} {\bibinfo {author} {\bibfnamefont {R.~F.}\ \bibnamefont
  {Sarkozy}}, \bibinfo {author} {\bibfnamefont {C.~W.}\ \bibnamefont
  {Moeller}}, \ and\ \bibinfo {author} {\bibfnamefont {B.~L.}\ \bibnamefont
  {Chamberland}},\ }\href@noop {} {\bibfield  {journal} {\bibinfo  {journal}
  {J. Solid State Chem.}\ }\textbf {\bibinfo {volume} {9}},\ \bibinfo {pages}
  {242} (\bibinfo {year} {1974})}\BibitemShut {NoStop}%
\bibitem [{\citenamefont {Dijksma}\ \emph {et~al.}(1993)\citenamefont
  {Dijksma}, \citenamefont {Vente}, \citenamefont {Frikkee},\ and\
  \citenamefont {IJdo}}]{Dijksma1993}%
  \BibitemOpen
  \bibfield  {author} {\bibinfo {author} {\bibfnamefont {F.~J.~J.}\
  \bibnamefont {Dijksma}}, \bibinfo {author} {\bibfnamefont {J.~F.}\
  \bibnamefont {Vente}}, \bibinfo {author} {\bibfnamefont {E.}~\bibnamefont
  {Frikkee}}, \ and\ \bibinfo {author} {\bibfnamefont {D.~J.~W.}\ \bibnamefont
  {IJdo}},\ }\href@noop {} {\bibfield  {journal} {\bibinfo  {journal} {Mat.
  Res. Bull.}\ }\textbf {\bibinfo {volume} {28}},\ \bibinfo {pages} {1145}
  (\bibinfo {year} {1993})}\BibitemShut {NoStop}%
\bibitem [{\citenamefont {Cao}\ \emph {et~al.}(2007)\citenamefont {Cao},
  \citenamefont {Durairaj}, \citenamefont {Chikara}, \citenamefont {Parkin},\
  and\ \citenamefont {Schlottmann}}]{Cao2007}%
  \BibitemOpen
  \bibfield  {author} {\bibinfo {author} {\bibfnamefont {G.}~\bibnamefont
  {Cao}}, \bibinfo {author} {\bibfnamefont {V.}~\bibnamefont {Durairaj}},
  \bibinfo {author} {\bibfnamefont {S.}~\bibnamefont {Chikara}}, \bibinfo
  {author} {\bibfnamefont {S.}~\bibnamefont {Parkin}}, \ and\ \bibinfo {author}
  {\bibfnamefont {P.}~\bibnamefont {Schlottmann}},\ }\href {\doibase
  10.1103/PhysRevB.75.134402} {\bibfield  {journal} {\bibinfo  {journal} {Phys.
  Rev. B}\ }\textbf {\bibinfo {volume} {75}},\ \bibinfo {pages} {134402}
  (\bibinfo {year} {2007})}\BibitemShut {NoStop}%
\bibitem [{\citenamefont {D\"oring}\ \emph {et~al.}(2004)\citenamefont
  {D\"oring}, \citenamefont {Sternemann}, \citenamefont {Kaprolat},
  \citenamefont {Mattila}, \citenamefont {H\"am\"al\"ainen},\ and\
  \citenamefont {Sch\"ulke}}]{Doring2004}%
  \BibitemOpen
  \bibfield  {author} {\bibinfo {author} {\bibfnamefont {G.}~\bibnamefont
  {D\"oring}}, \bibinfo {author} {\bibfnamefont {C.}~\bibnamefont
  {Sternemann}}, \bibinfo {author} {\bibfnamefont {A.}~\bibnamefont
  {Kaprolat}}, \bibinfo {author} {\bibfnamefont {A.}~\bibnamefont {Mattila}},
  \bibinfo {author} {\bibfnamefont {K.}~\bibnamefont {H\"am\"al\"ainen}}, \
  and\ \bibinfo {author} {\bibfnamefont {W.}~\bibnamefont {Sch\"ulke}},\ }\href
  {\doibase 10.1103/PhysRevB.70.085115} {\bibfield  {journal} {\bibinfo
  {journal} {Phys. Rev. B}\ }\textbf {\bibinfo {volume} {70}},\ \bibinfo
  {pages} {085115} (\bibinfo {year} {2004})}\BibitemShut {NoStop}%
\bibitem [{\citenamefont {Krause}\ and\ \citenamefont
  {Oliver}(1979)}]{Krause1979}%
  \BibitemOpen
  \bibfield  {author} {\bibinfo {author} {\bibfnamefont {M.~O.}\ \bibnamefont
  {Krause}}\ and\ \bibinfo {author} {\bibfnamefont {J.~H.}\ \bibnamefont
  {Oliver}},\ }\href@noop {} {\bibfield  {journal} {\bibinfo  {journal} {J.
  Phys. Chem. Ref. Data}\ }\textbf {\bibinfo {volume} {8}},\ \bibinfo {pages}
  {329} (\bibinfo {year} {1979})}\BibitemShut {NoStop}%
\bibitem [{\citenamefont {Clancy}\ \emph {et~al.}(2012)\citenamefont {Clancy},
  \citenamefont {Chen}, \citenamefont {Kim}, \citenamefont {Chen},
  \citenamefont {Plumb}, \citenamefont {Jeon}, \citenamefont {Noh},\ and\
  \citenamefont {Kim}}]{Clancy2012}%
  \BibitemOpen
  \bibfield  {author} {\bibinfo {author} {\bibfnamefont {J.~P.}\ \bibnamefont
  {Clancy}}, \bibinfo {author} {\bibfnamefont {N.}~\bibnamefont {Chen}},
  \bibinfo {author} {\bibfnamefont {C.~Y.}\ \bibnamefont {Kim}}, \bibinfo
  {author} {\bibfnamefont {W.~F.}\ \bibnamefont {Chen}}, \bibinfo {author}
  {\bibfnamefont {K.~W.}\ \bibnamefont {Plumb}}, \bibinfo {author}
  {\bibfnamefont {B.~C.}\ \bibnamefont {Jeon}}, \bibinfo {author}
  {\bibfnamefont {T.~W.}\ \bibnamefont {Noh}}, \ and\ \bibinfo {author}
  {\bibfnamefont {Y.-J.}\ \bibnamefont {Kim}},\ }\href {\doibase
  10.1103/PhysRevB.86.195131} {\bibfield  {journal} {\bibinfo  {journal} {Phys.
  Rev. B}\ }\textbf {\bibinfo {volume} {86}},\ \bibinfo {pages} {195131}
  (\bibinfo {year} {2012})}\BibitemShut {NoStop}%
\bibitem [{\citenamefont {H\"am\"al\"ainen}\ \emph {et~al.}(1991)\citenamefont
  {H\"am\"al\"ainen}, \citenamefont {Siddons}, \citenamefont {Hastings},\ and\
  \citenamefont {Berman}}]{Hamalainen1991}%
  \BibitemOpen
  \bibfield  {author} {\bibinfo {author} {\bibfnamefont {K.}~\bibnamefont
  {H\"am\"al\"ainen}}, \bibinfo {author} {\bibfnamefont {D.~P.}\ \bibnamefont
  {Siddons}}, \bibinfo {author} {\bibfnamefont {J.~B.}\ \bibnamefont
  {Hastings}}, \ and\ \bibinfo {author} {\bibfnamefont {L.~E.}\ \bibnamefont
  {Berman}},\ }\href {\doibase 10.1103/PhysRevLett.67.2850} {\bibfield
  {journal} {\bibinfo  {journal} {Phys. Rev. Lett.}\ }\textbf {\bibinfo
  {volume} {67}},\ \bibinfo {pages} {2850} (\bibinfo {year}
  {1991})}\BibitemShut {NoStop}%
\bibitem [{\citenamefont {van~der Laan}\ and\ \citenamefont
  {Kirkman}(1992)}]{Laan1992}%
  \BibitemOpen
  \bibfield  {author} {\bibinfo {author} {\bibfnamefont {G.}~\bibnamefont
  {van~der Laan}}\ and\ \bibinfo {author} {\bibfnamefont {I.~W.}\ \bibnamefont
  {Kirkman}},\ }\href {http://stacks.iop.org/0953-8984/4/i=16/a=019} {\bibfield
   {journal} {\bibinfo  {journal} {Journal of Physics: Condensed Matter}\
  }\textbf {\bibinfo {volume} {4}},\ \bibinfo {pages} {4189} (\bibinfo {year}
  {1992})}\BibitemShut {NoStop}%
\bibitem [{\citenamefont {Laguna-Marco}\ \emph {et~al.}(2015)\citenamefont
  {Laguna-Marco}, \citenamefont {Kayser}, \citenamefont {Alonso}, \citenamefont
  {Mart\'{\i}nez-Lope}, \citenamefont {van Veenendaal}, \citenamefont {Choi},\
  and\ \citenamefont {Haskel}}]{LagunaMarco2015}%
  \BibitemOpen
  \bibfield  {author} {\bibinfo {author} {\bibfnamefont {M.~A.}\ \bibnamefont
  {Laguna-Marco}}, \bibinfo {author} {\bibfnamefont {P.}~\bibnamefont
  {Kayser}}, \bibinfo {author} {\bibfnamefont {J.~A.}\ \bibnamefont {Alonso}},
  \bibinfo {author} {\bibfnamefont {M.~J.}\ \bibnamefont {Mart\'{\i}nez-Lope}},
  \bibinfo {author} {\bibfnamefont {M.}~\bibnamefont {van Veenendaal}},
  \bibinfo {author} {\bibfnamefont {Y.}~\bibnamefont {Choi}}, \ and\ \bibinfo
  {author} {\bibfnamefont {D.}~\bibnamefont {Haskel}},\ }\href {\doibase
  10.1103/PhysRevB.91.214433} {\bibfield  {journal} {\bibinfo  {journal} {Phys.
  Rev. B}\ }\textbf {\bibinfo {volume} {91}},\ \bibinfo {pages} {214433}
  (\bibinfo {year} {2015})}\BibitemShut {NoStop}%
\bibitem [{Note1()}]{Note1}%
  \BibitemOpen
  \bibinfo {note} {True $M_4$ XAS final state would have a hole in $3d_{3/2}$
  and an extra electron in $6p-5f$ states instead of $5d$ states.}\BibitemShut
  {Stop}%
\bibitem [{\citenamefont {Liu}\ \emph {et~al.}(2015)\citenamefont {Liu},
  \citenamefont {Dean}, \citenamefont {Liu}, \citenamefont {Chiuzbăian},
  \citenamefont {Jaouen}, \citenamefont {Nicolaou}, \citenamefont {Yin},
  \citenamefont {Serrao}, \citenamefont {Ramesh}, \citenamefont {Ding},\ and\
  \citenamefont {Hill}}]{Liu2015}%
  \BibitemOpen
  \bibfield  {author} {\bibinfo {author} {\bibfnamefont {X.}~\bibnamefont
  {Liu}}, \bibinfo {author} {\bibfnamefont {M.~P.~M.}\ \bibnamefont {Dean}},
  \bibinfo {author} {\bibfnamefont {J.}~\bibnamefont {Liu}}, \bibinfo {author}
  {\bibfnamefont {S.~G.}\ \bibnamefont {Chiuzbăian}}, \bibinfo {author}
  {\bibfnamefont {N.}~\bibnamefont {Jaouen}}, \bibinfo {author} {\bibfnamefont
  {A.}~\bibnamefont {Nicolaou}}, \bibinfo {author} {\bibfnamefont {W.~G.}\
  \bibnamefont {Yin}}, \bibinfo {author} {\bibfnamefont {C.~R.}\ \bibnamefont
  {Serrao}}, \bibinfo {author} {\bibfnamefont {R.}~\bibnamefont {Ramesh}},
  \bibinfo {author} {\bibfnamefont {H.}~\bibnamefont {Ding}}, \ and\ \bibinfo
  {author} {\bibfnamefont {J.~P.}\ \bibnamefont {Hill}},\ }\href
  {http://stacks.iop.org/0953-8984/27/i=20/a=202202} {\bibfield  {journal}
  {\bibinfo  {journal} {Journal of Physics: Condensed Matter}\ }\textbf
  {\bibinfo {volume} {27}},\ \bibinfo {pages} {202202} (\bibinfo {year}
  {2015})}\BibitemShut {NoStop}%
\bibitem [{\citenamefont {Kim}\ \emph {et~al.}(2002)\citenamefont {Kim},
  \citenamefont {Hill}, \citenamefont {Burns}, \citenamefont {Wakimoto},
  \citenamefont {Birgeneau}, \citenamefont {Casa}, \citenamefont {Gog},\ and\
  \citenamefont {Venkataraman}}]{LCO}%
  \BibitemOpen
  \bibfield  {author} {\bibinfo {author} {\bibfnamefont {Y.-J.}\ \bibnamefont
  {Kim}}, \bibinfo {author} {\bibfnamefont {J.}~\bibnamefont {Hill}}, \bibinfo
  {author} {\bibfnamefont {C.}~\bibnamefont {Burns}}, \bibinfo {author}
  {\bibfnamefont {S.}~\bibnamefont {Wakimoto}}, \bibinfo {author}
  {\bibfnamefont {R.}~\bibnamefont {Birgeneau}}, \bibinfo {author}
  {\bibfnamefont {D.}~\bibnamefont {Casa}}, \bibinfo {author} {\bibfnamefont
  {T.}~\bibnamefont {Gog}}, \ and\ \bibinfo {author} {\bibfnamefont
  {C.}~\bibnamefont {Venkataraman}},\ }\href@noop {} {\bibfield  {journal}
  {\bibinfo  {journal} {Physical Review Letters}\ }\textbf {\bibinfo {volume}
  {89}},\ \bibinfo {pages} {177003} (\bibinfo {year} {2002})}\BibitemShut
  {NoStop}%
\bibitem [{\citenamefont {Lu}\ \emph {et~al.}(2005)\citenamefont {Lu},
  \citenamefont {Chabot-Couture}, \citenamefont {Zhao}, \citenamefont
  {Hancock}, \citenamefont {Kaneko}, \citenamefont {Vajk}, \citenamefont {Yu},
  \citenamefont {Grenier}, \citenamefont {Kim}, \citenamefont {Casa},
  \citenamefont {Gog},\ and\ \citenamefont {Greven}}]{Lu2005}%
  \BibitemOpen
  \bibfield  {author} {\bibinfo {author} {\bibfnamefont {L.}~\bibnamefont
  {Lu}}, \bibinfo {author} {\bibfnamefont {G.}~\bibnamefont {Chabot-Couture}},
  \bibinfo {author} {\bibfnamefont {X.}~\bibnamefont {Zhao}}, \bibinfo {author}
  {\bibfnamefont {J.}~\bibnamefont {Hancock}}, \bibinfo {author} {\bibfnamefont
  {N.}~\bibnamefont {Kaneko}}, \bibinfo {author} {\bibfnamefont
  {O.}~\bibnamefont {Vajk}}, \bibinfo {author} {\bibfnamefont {G.}~\bibnamefont
  {Yu}}, \bibinfo {author} {\bibfnamefont {S.}~\bibnamefont {Grenier}},
  \bibinfo {author} {\bibfnamefont {Y.}~\bibnamefont {Kim}}, \bibinfo {author}
  {\bibfnamefont {D.}~\bibnamefont {Casa}}, \bibinfo {author} {\bibfnamefont
  {T.}~\bibnamefont {Gog}}, \ and\ \bibinfo {author} {\bibfnamefont
  {M.}~\bibnamefont {Greven}},\ }\href {\doibase 10.1103/PhysRevLett.95.217003}
  {\bibfield  {journal} {\bibinfo  {journal} {Physical Review Letters}\
  }\textbf {\bibinfo {volume} {95}},\ \bibinfo {pages} {217003} (\bibinfo
  {year} {2005})}\BibitemShut {NoStop}%
\bibitem [{\citenamefont {Kim}\ \emph {et~al.}(2007)\citenamefont {Kim},
  \citenamefont {Hill}, \citenamefont {Wakimoto}, \citenamefont {Birgeneau},
  \citenamefont {Chou}, \citenamefont {Motoyama}, \citenamefont {Kojima},
  \citenamefont {Uchida}, \citenamefont {Casa},\ and\ \citenamefont
  {Gog}}]{Observations}%
  \BibitemOpen
  \bibfield  {author} {\bibinfo {author} {\bibfnamefont {Y.-J.}\ \bibnamefont
  {Kim}}, \bibinfo {author} {\bibfnamefont {J.~P.}\ \bibnamefont {Hill}},
  \bibinfo {author} {\bibfnamefont {S.}~\bibnamefont {Wakimoto}}, \bibinfo
  {author} {\bibfnamefont {R.~J.}\ \bibnamefont {Birgeneau}}, \bibinfo {author}
  {\bibfnamefont {F.~C.}\ \bibnamefont {Chou}}, \bibinfo {author}
  {\bibfnamefont {N.}~\bibnamefont {Motoyama}}, \bibinfo {author}
  {\bibfnamefont {K.~M.}\ \bibnamefont {Kojima}}, \bibinfo {author}
  {\bibfnamefont {S.}~\bibnamefont {Uchida}}, \bibinfo {author} {\bibfnamefont
  {D.}~\bibnamefont {Casa}}, \ and\ \bibinfo {author} {\bibfnamefont
  {T.}~\bibnamefont {Gog}},\ }\href {\doibase 10.1103/PhysRevB.76.155116}
  {\bibfield  {journal} {\bibinfo  {journal} {Physical Review B}\ }\textbf
  {\bibinfo {volume} {76}},\ \bibinfo {pages} {155116} (\bibinfo {year}
  {2007})}\BibitemShut {NoStop}%
\bibitem [{\citenamefont {van~den Brink}\ and\ \citenamefont {van
  Veenendaal}(2006)}]{UCL-EPL}%
  \BibitemOpen
  \bibfield  {author} {\bibinfo {author} {\bibfnamefont {J.}~\bibnamefont
  {van~den Brink}}\ and\ \bibinfo {author} {\bibfnamefont {M.}~\bibnamefont
  {van Veenendaal}},\ }\href {http://stacks.iop.org/0295-5075/73/i=1/a=121}
  {\bibfield  {journal} {\bibinfo  {journal} {EPL (Europhysics Letters)}\
  }\textbf {\bibinfo {volume} {73}},\ \bibinfo {pages} {121} (\bibinfo {year}
  {2006})}\BibitemShut {NoStop}%
\bibitem [{\citenamefont {Magnuson}\ \emph {et~al.}(2003)\citenamefont
  {Magnuson}, \citenamefont {Rubensson}, \citenamefont {F\"ohlisch},
  \citenamefont {Wassdahl}, \citenamefont {Nilsson},\ and\ \citenamefont
  {M\aa{}rtensson}}]{Magnuson2003}%
  \BibitemOpen
  \bibfield  {author} {\bibinfo {author} {\bibfnamefont {M.}~\bibnamefont
  {Magnuson}}, \bibinfo {author} {\bibfnamefont {J.-E.}\ \bibnamefont
  {Rubensson}}, \bibinfo {author} {\bibfnamefont {A.}~\bibnamefont
  {F\"ohlisch}}, \bibinfo {author} {\bibfnamefont {N.}~\bibnamefont
  {Wassdahl}}, \bibinfo {author} {\bibfnamefont {A.}~\bibnamefont {Nilsson}}, \
  and\ \bibinfo {author} {\bibfnamefont {N.}~\bibnamefont {M\aa{}rtensson}},\
  }\href {\doibase 10.1103/PhysRevB.68.045119} {\bibfield  {journal} {\bibinfo
  {journal} {Phys. Rev. B}\ }\textbf {\bibinfo {volume} {68}},\ \bibinfo
  {pages} {045119} (\bibinfo {year} {2003})}\BibitemShut {NoStop}%
\bibitem [{\citenamefont {Kahk}\ \emph {et~al.}(2014)\citenamefont {Kahk},
  \citenamefont {Poll}, \citenamefont {Oropeza}, \citenamefont {Ablett},
  \citenamefont {C\'eolin}, \citenamefont {Rueff}, \citenamefont {Agrestini},
  \citenamefont {Utsumi}, \citenamefont {Tsuei}, \citenamefont {Liao},
  \citenamefont {Borgatti}, \citenamefont {Panaccione}, \citenamefont
  {Regoutz}, \citenamefont {Egdell}, \citenamefont {Morgan}, \citenamefont
  {Scanlon},\ and\ \citenamefont {Payne}}]{Kahk2014}%
  \BibitemOpen
  \bibfield  {author} {\bibinfo {author} {\bibfnamefont {J.~M.}\ \bibnamefont
  {Kahk}}, \bibinfo {author} {\bibfnamefont {C.~G.}\ \bibnamefont {Poll}},
  \bibinfo {author} {\bibfnamefont {F.~E.}\ \bibnamefont {Oropeza}}, \bibinfo
  {author} {\bibfnamefont {J.~M.}\ \bibnamefont {Ablett}}, \bibinfo {author}
  {\bibfnamefont {D.}~\bibnamefont {C\'eolin}}, \bibinfo {author}
  {\bibfnamefont {J.-P.}\ \bibnamefont {Rueff}}, \bibinfo {author}
  {\bibfnamefont {S.}~\bibnamefont {Agrestini}}, \bibinfo {author}
  {\bibfnamefont {Y.}~\bibnamefont {Utsumi}}, \bibinfo {author} {\bibfnamefont
  {K.~D.}\ \bibnamefont {Tsuei}}, \bibinfo {author} {\bibfnamefont {Y.~F.}\
  \bibnamefont {Liao}}, \bibinfo {author} {\bibfnamefont {F.}~\bibnamefont
  {Borgatti}}, \bibinfo {author} {\bibfnamefont {G.}~\bibnamefont
  {Panaccione}}, \bibinfo {author} {\bibfnamefont {A.}~\bibnamefont {Regoutz}},
  \bibinfo {author} {\bibfnamefont {R.~G.}\ \bibnamefont {Egdell}}, \bibinfo
  {author} {\bibfnamefont {B.~J.}\ \bibnamefont {Morgan}}, \bibinfo {author}
  {\bibfnamefont {D.~O.}\ \bibnamefont {Scanlon}}, \ and\ \bibinfo {author}
  {\bibfnamefont {D.~J.}\ \bibnamefont {Payne}},\ }\href {\doibase
  10.1103/PhysRevLett.112.117601} {\bibfield  {journal} {\bibinfo  {journal}
  {Phys. Rev. Lett.}\ }\textbf {\bibinfo {volume} {112}},\ \bibinfo {pages}
  {117601} (\bibinfo {year} {2014})}\BibitemShut {NoStop}%
\end{thebibliography}
\end{document}